\documentclass{article} %
\usepackage{configuration/iclr2026_conference,times}

\usepackage{hyperref}
\usepackage{url}
\usepackage{enumitem}

\usepackage{booktabs}
\usepackage{xcolor} %
\usepackage{graphicx} %
\usepackage{multirow}
\usepackage{array}
\usepackage{makecell}

\usepackage{amsmath,amssymb,amsthm}

\newenvironment{talign*}
{\csname align*\endcsname}
{\endalign}

\usepackage[utf8]{inputenc}         %
\usepackage[T1]{fontenc}            %
\usepackage{url}                    %
\usepackage{booktabs}               %
\usepackage{amsfonts}               %
\usepackage{nicefrac}               %
\usepackage{microtype}              %
\usepackage{algorithm}
\usepackage{algpseudocode}  %
\algnewcommand\algorithmicinput{\textbf{Input:}}
\algnewcommand\Input{\item[\algorithmicinput]}

\usepackage{graphicx}
\usepackage{subcaption}
\usepackage[flushleft]{threeparttable}
\usepackage{float}
\usepackage{multirow}
\usepackage{xspace}
\usepackage{enumitem}
\usepackage[font=small]{caption}
\usepackage{autobreak}
\usepackage{sidecap}
\usepackage{wrapfig}
\usepackage{bbding}
\usepackage[toc, page, header]{appendix}
\usepackage{tikz}
\usepackage{pifont}
\usepackage{mdframed}
\usepackage{colortbl}

\usepackage{iclr2026_conference,times}
\usepackage[most]{tcolorbox}
\tcbuselibrary{listingsutf8}
\usepackage{mdframed}
\usepackage{hyperref}
\usepackage{url}
\usepackage{booktabs}
\usepackage{xcolor} %

\usepackage{graphicx} %
\usepackage{multirow}
\usepackage{booktabs}
\usepackage{array}
\usepackage{enumitem}
\usepackage{makecell}

\hypersetup{
    colorlinks=true,
    linkcolor=blue,
    citecolor=blue,
    urlcolor=blue
}

\definecolor{coral}{RGB}{255,127,80}
\definecolor{darkgreen}{RGB}{0,100,0}
\definecolor{darkyellow}{RGB}{204,153,0}
\definecolor{salmon}{RGB}{250,128,114}
\definecolor{darkred}{RGB}{150,0,0}
\newcommand{\darkredtext}[1]{{\color{darkred}#1}}

\newcommand{\secref}[1]{\hyperref[#1]{\darkredtext{Sec.~\ref*{#1}}}}
\newcommand{\thmref}[1]{\hyperref[#1]{\darkredtext{Thm.~\ref*{#1}}}}
\newcommand{\defref}[1]{\hyperref[#1]{\darkredtext{Def.~\ref*{#1}}}}
\newcommand{\propref}[1]{\hyperref[#1]{\darkredtext{Prop.~\ref*{#1}}}}
\newcommand{\assumpref}[1]{\hyperref[#1]{\darkredtext{Assump.~\ref*{#1}}}}
\newcommand{\remarkref}[1]{\hyperref[#1]{\darkredtext{Rem.~\ref*{#1}}}}
\newcommand{\hypref}[1]{\hyperref[#1]{\darkredtext{Hyp.~\ref*{#1}}}}
\newcommand{\conjref}[1]{\hyperref[#1]{\darkredtext{Conj.~\ref*{#1}}}}
\newcommand{\lemref}[1]{\hyperref[#1]{\darkredtext{Lem.~\ref*{#1}}}}
\newcommand{\corref}[1]{\hyperref[#1]{\darkredtext{Cor.~\ref*{#1}}}}
\newcommand{\noteref}[1]{\hyperref[#1]{\darkredtext{Nota.~\ref*{#1}}}}
\newcommand{\claimref}[1]{\hyperref[#1]{\darkredtext{Clm.~\ref*{#1}}}}
\newcommand{\obsref}[1]{\hyperref[#1]{\darkredtext{Obs.~\ref*{#1}}}}
\newcommand{\myalgref}[1]{\hyperref[#1]{\darkredtext{Alg.~\ref*{#1}}}}
\newcommand{\figref}[1]{\hyperref[#1]{\darkredtext{Fig.~\ref*{#1}}}}
\newcommand{\tabref}[1]{\hyperref[#1]{\darkredtext{Tab.~\ref*{#1}}}}
\newcommand{\appref}[1]{\hyperref[#1]{\darkredtext{App.~\ref*{#1}}}}

\newtheoremstyle{custom}
{1pt} %
{1pt} %
{\itshape} %
{} %
{\bfseries} %
{} %
{ } %
{\thmname{#1} \thmnumber{#2} \thmnote{(#3)} . } %

\theoremstyle{custom}

\newtheorem{innerdefinition}{Definition}
\newtheorem{innerproposition}{Proposition}
\newtheorem{innerassumption}{Assumption}
\newtheorem{innerremark}{Remark}
\newtheorem{innertheorem}{Theorem}
\newtheorem{innerhypothesis}{Hypothesis}
\newtheorem{innerconjecture}{Conjecture}
\newtheorem{innerlemma}{Lemma}
\newtheorem{innercorollary}{Corollary}
\newtheorem{innernotation}{Notation}
\newtheorem{innerclaim}{Claim}
\newtheorem{innerproblem}{Problem}

\newtheorem{innerobservation}{Observation}

\newmdenv[
    backgroundcolor=gray!10,
    linecolor=gray!100,
    linewidth=0.8pt,
    skipabove=2pt,
    skipbelow=2pt,
    innertopmargin=10pt,
    innerbottommargin=5pt,
    innerleftmargin=5pt,
    innerrightmargin=5pt,
]{definitionframe}

\newmdenv[
    backgroundcolor=blue!10,
    linecolor=blue!100,
    linewidth=0.8pt,
    skipabove=2pt,
    skipbelow=2pt,
    innertopmargin=10pt,
    innerbottommargin=5pt,
    innerleftmargin=5pt,
    innerrightmargin=5pt,
]{propositionframe}

\newmdenv[
    backgroundcolor=green!10,
    linecolor=green!100,
    linewidth=0.8pt,
    skipabove=2pt,
    skipbelow=2pt,
    innertopmargin=10pt,
    innerbottommargin=5pt,
    innerleftmargin=5pt,
    innerrightmargin=5pt,
]{assumptionframe}

\newmdenv[
    backgroundcolor=yellow!10,
    linecolor=yellow!100,
    linewidth=0.8pt,
    skipabove=2pt,
    skipbelow=2pt,
    innertopmargin=10pt,
    innerbottommargin=5pt,
    innerleftmargin=5pt,
    innerrightmargin=5pt,
]{remarkframe}

\newmdenv[
    backgroundcolor=red!10,
    linecolor=red!100,
    linewidth=0.8pt,
    skipabove=2pt,
    skipbelow=2pt,
    innertopmargin=10pt,
    innerbottommargin=5pt,
    innerleftmargin=5pt,
    innerrightmargin=5pt,
]{theoremframe}

\newmdenv[
    backgroundcolor=purple!10,
    linecolor=purple!100,
    linewidth=0.8pt,
    skipabove=2pt,
    skipbelow=2pt,
    innertopmargin=10pt,
    innerbottommargin=5pt,
    innerleftmargin=5pt,
    innerrightmargin=5pt,
]{hypothesisframe}

\newmdenv[
    backgroundcolor=orange!10,
    linecolor=orange!100,
    linewidth=0.8pt,
    skipabove=2pt,
    skipbelow=2pt,
    innertopmargin=10pt,
    innerbottommargin=5pt,
    innerleftmargin=5pt,
    innerrightmargin=5pt,
]{conjectureframe}

\newmdenv[
    backgroundcolor=cyan!10,
    linecolor=cyan!100,
    linewidth=0.8pt,
    skipabove=2pt,
    skipbelow=2pt,
    innertopmargin=10pt,
    innerbottommargin=5pt,
    innerleftmargin=5pt,
    innerrightmargin=5pt,
]{lemmaframe}

\newmdenv[
    backgroundcolor=magenta!10,
    linecolor=magenta!100,
    linewidth=0.8pt,
    skipabove=2pt,
    skipbelow=2pt,
    innertopmargin=10pt,
    innerbottommargin=5pt,
    innerleftmargin=5pt,
    innerrightmargin=5pt,
]{corollaryframe}

\newmdenv[
    backgroundcolor=pink!10,
    linecolor=pink!100,
    linewidth=0.8pt,
    skipabove=2pt,
    skipbelow=2pt,
    innertopmargin=10pt,
    innerbottommargin=5pt,
    innerleftmargin=5pt,
    innerrightmargin=5pt,
]{notationframe}

\newmdenv[
    backgroundcolor=violet!10,
    linecolor=violet!100,
    linewidth=0.8pt,
    skipabove=2pt,
    skipbelow=2pt,
    innertopmargin=10pt,
    innerbottommargin=5pt,
    innerleftmargin=5pt,
    innerrightmargin=5pt,
]{claimframe}

\newmdenv[
    backgroundcolor=salmon!10,
    linecolor=salmon!100,
    linewidth=0.8pt,
    skipabove=2pt,
    skipbelow=2pt,
    innertopmargin=10pt,
    innerbottommargin=5pt,
    innerleftmargin=5pt,
    innerrightmargin=5pt,
]{problemframe}

\newmdenv[
    backgroundcolor=lavender!10,
    linecolor=lavender!100,
    linewidth=0.8pt,
    skipabove=2pt,
    skipbelow=2pt,
    innertopmargin=10pt,
    innerbottommargin=5pt,
    innerleftmargin=5pt,
    innerrightmargin=5pt,
]{observationframe}

\newenvironment{definition}
{\begin{definitionframe}\begin{innerdefinition}}
            {\end{innerdefinition}\end{definitionframe}}

\newcommand{\method}{\textsc{SupervisorAgent}\xspace} %

\usepackage[textwidth=3.5cm]{todonotes}

\usepackage{newunicodechar}
\newunicodechar{↑}{$\uparrow$}
\newunicodechar{↓}{$\downarrow$}

\title{Stop Wasting Your Tokens: Towards Efficient Runtime Multi-Agent Systems}

\author{%
    Fulin Lin$^{1,}$\thanks{Work was done during Fulin's visit to Westlake University.} \quad Shaowen Chen$^{1}$ \quad Ruishan Fang$^{2,1}$ \quad Hongwei Wang$^{1,3,\dag}$  \quad Tao Lin$^{2,}$\thanks{Corresponding authors.}\\
  $^1$Zhejiang University \quad
  $^2$Westlake University \\
  $^3$State Key Laboratory of CAD\&CG, Zhejiang University \\
  \texttt{\{fulin1.24, hongweiwang\}@intl.zju.edu.cn} \quad \texttt{swenchen@zju.edu.cn} \\ \texttt{\{fangruishan, lintao\}@westlake.edu.cn}
}

\iclrfinalcopy %
\begin{document}

\maketitle

\vspace{-1.5em}

\begin{abstract}
    While Multi-Agent Systems (MAS) excel at complex tasks, their growing autonomy with operational complexity often leads to critical inefficiencies, such as excessive token consumption and failures arising from misinformation.
    Existing methods primarily focus on post-hoc failure attribution, lacking proactive, real-time interventions to enhance robustness and efficiency.
    To this end, we introduce \method, a lightweight and modular framework for runtime, adaptive supervision that operates without altering the base agent's architecture.
    Triggered by an LLM-free adaptive filter, \method intervenes at critical junctures to proactively correct errors, guide inefficient behaviors, and purify observations.
    On the challenging GAIA benchmark, \method reduces the token consumption of the Smolagent framework by an average of 29.68\% without compromising its success rate.
    Extensive experiments across five additional benchmarks (math reasoning, code generation, and question answering) and various SoTA foundation models validate the broad applicability and robustness of our approach.
    \looseness=-1
\end{abstract}
\vspace{-1em}

\section{Introduction}

The advent of powerful Large Language Models (LLMs) has catalyzed significant advancements in Multi-Agent Systems (MAS)~\citep{liu2025advanceschallengesfoundationagents,gao2025surveyselfevolvingagentspath}, enabling them to achieve remarkable performance across diverse and challenging domains such as mathematical reasoning~\citep{shang2025rstar2agentagenticreasoningtechnical}, code generation~\citep{lu2025requirementsdevelopmentformalizationreliable}, and complex question answering~\citep{luo2025entitylinkingagentquestion}.
This progress has spurred research into sophisticated agent architectures, including self-evolving systems that learn from feedback and experience~\citep{shi2025mobileguirladvancingmobilegui,liu2025infiguir1advancingmultimodalgui}, and dynamic topologies that adapt to task complexity~\citep{li2025assemblecrewautomaticmultiagent,li2025chainofagentsendtoendagentfoundation}.
However, a critical paradox has emerged: as these systems grow more capable and complex, they often become less robust and economically viable~\citep{wu2025dissectingadversarialrobustnessmultimodal, huang2025competing}.
Systemic inefficiencies incur prohibitive computational costs, while intricate interactions introduce vectors for unpredictable failures~\citep{zhang2025agentcausestaskfailures}.
\looseness=-1

\begin{figure}[!t] %
    \centering
    \includegraphics[width=0.9\linewidth]{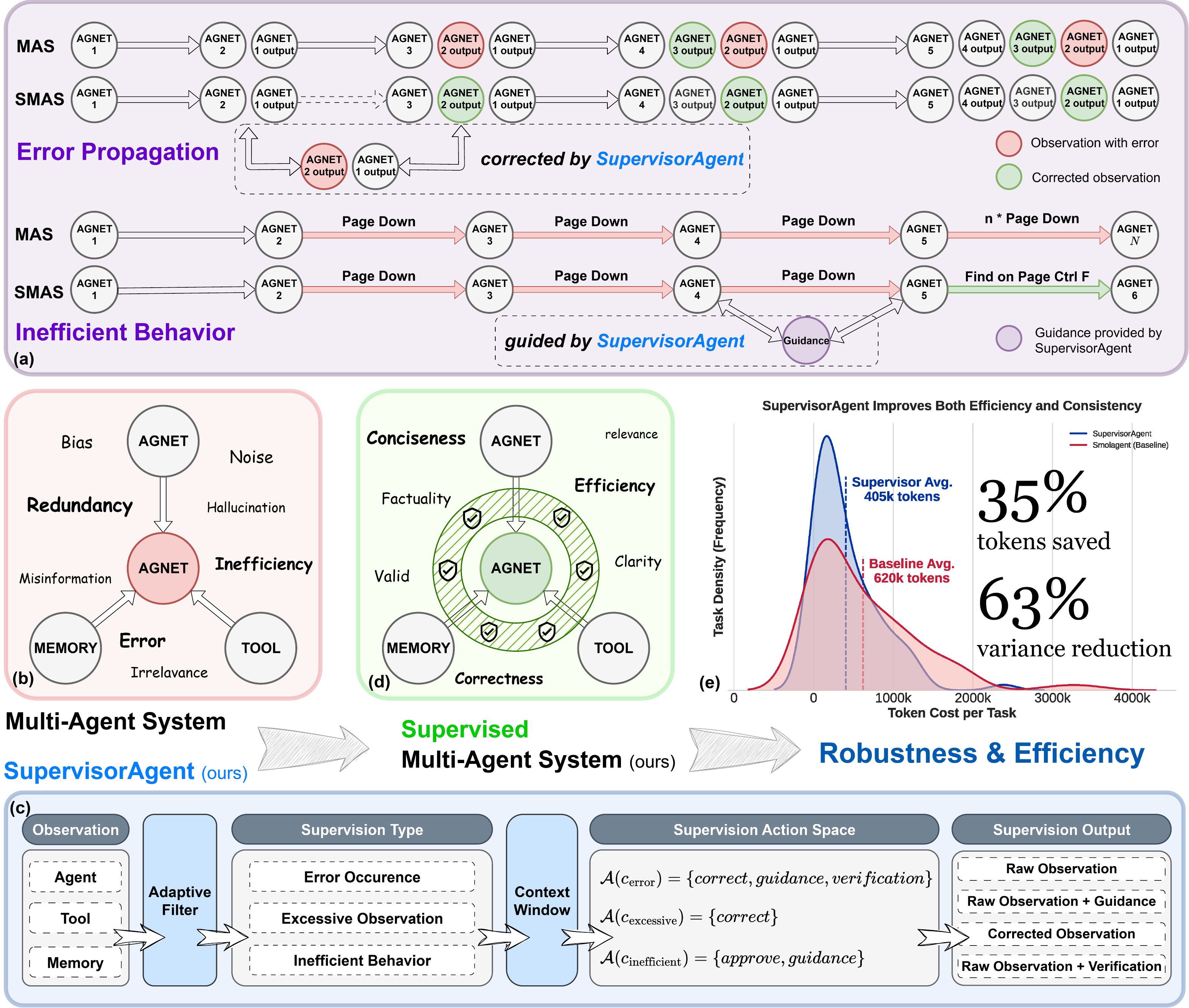}
    \caption{\small
        \textbf{The \method Framework: Concept and Impact.}
        \textbf{(a)}~Illustrative examples of common failure modes in MAS, including \textbf{error propagation} and \textbf{inefficient loops}, and the corresponding intervention by our \method.
        \textbf{(b)}~An overview of a conventional MAS, highlighting the high-risk interaction loci (agent-agent, agent-tool, agent-memory) where such failures occur.
        \textbf{(c)}~The core workflow of our \method, which monitors these interactions to provide real-time intervention.
        \textbf{(d)}~The resulting Supervised MAS (SMAS), which integrates the \method to enhance robustness and efficiency.
        \textbf{(e)}~Performance on GAIA (Level 2), where SMAS (blue) reduces token cost by 35\% and variance by 63\% versus the baseline (red).
    }
    \vspace{-2em}
    \label{fig:introduction}
\end{figure}

This lack of robustness stems from the operational complexity of modern MAS, which introduces a significant reliability challenge~\citep{tian2025outlookopportunitieschallengesmultiagent}.
The long chain of interactions inherent in these systems creates fertile ground for \textbf{error propagation}~\citep{dong2025practicalmemoryinjectionattack,shen2025understandinginformationpropagationeffects}.
For instance, a single piece of misinformation generated by an agent, a common risk with today's powerful yet occasionally hallucinatory foundation models~\citep{kalai2025languagemodelshallucinate,farquhar_detecting_2024}, can be committed to memory and subsequently poison the reasoning of all downstream agents (as explained in Figure \ref{fig:introduction}a).
These vulnerabilities mean that even a state-of-the-art MAS can fail on tasks well within its theoretical capabilities, simply due to a lack of operational robustness~\citep{chen2024agentpoisonredteamingllmagents}.
\looseness=-1

Furthermore, the issue of \textbf{economic inefficiency} is a major barrier to the real-world deployment of MAS~\citep{wang2025efficientagentsbuildingeffective}.
We identify two primary sources of this inefficiency.
First, agents often struggle with long observations, such as verbose web pages or tool outputs, which flood their context windows.
This not only inflates token costs but can also obscure critical information, causing the agent to lose focus and derail its task execution~\citep{hosseini-etal-2025-efficient}.
Second, agents may adopt sub-optimal strategies, entering into repetitive action loops or choosing unnecessarily complex paths to a solution~\citep{cemri2025multiagentllmsystemsfail}, further wasting computational resources (see Figure \ref{fig:introduction}a).

To address these intertwined challenges, we propose \method, a lightweight and modular framework that enhances the \textbf{robustness} and \textbf{efficiency} of Multi-Agent Systems (MAS) through real-time supervision (see Figure \ref{fig:introduction}c).
Incorporating an adaptive filter, \method enables proactive process control, exemplified by its GAIA Level 2 performance in Figure~\ref{fig:introduction}e.
It adaptively intervenes at critical junctures to mitigate key operational risks: it conducts proactive error diagnosis, provides pragmatic guidance for inefficient behaviors, and performs adaptive observation purification to reduce contextual noise from long observations.

\textbf{In summary, our main contributions are:}
\begin{enumerate}[nosep, leftmargin=12pt]
    \item We propose and implement \textbf{\method}, a novel, lightweight, and non-intrusive meta-agent framework for real-time MAS supervision.
          It improves agent robustness and efficiency through proactive error correction, inefficiency guidance, and adaptive observation purification, without altering the base agents' architecture.
    \item We conduct extensive experiments on the challenging \textbf{GAIA} benchmark and demonstrate a significant \textbf{Pareto improvement}.
          When applied to the Smolagent framework~\citep{smolagents}, \method reduces token consumption by an average of \textbf{29.68\%} while maintaining competitive task success rates.
    \item We validate the \textbf{general applicability} of our approach across five additional benchmarks spanning mathematical reasoning, code generation, and question answering.
          Our method consistently delivers substantial efficiency gains, highlighted by a \textbf{23.74\%} token reduction on HumanEval alongside an accuracy improvement.
          The framework's effectiveness is further confirmed across various foundation models, including the GPT-4.1, Gemini-2.5-pro, and Qwen3 series.
\end{enumerate}

\section{Related Work}

\paragraph{The increasing complexity of Multi-Agent Systems (MAS).}
Recent advancements in Large Language Models have spurred the development of increasingly sophisticated Multi-Agent Systems (MAS) capable of tackling complex, multi-step tasks~\citep{tran2025multiagentcollaborationmechanismssurvey,he2025llm}.
Frameworks like Tongyi DeepResearch~\citep{tongyidr}, AgentOrchestra~\citep{zhang2025agentorchestrahierarchicalmultiagentframework}, and Aime~\citep{shi2025aimefullyautonomousmultiagentframework} exemplify this trend, introducing complex features such as hierarchical structures~\citep{zhu2025oagentsempiricalstudybuilding,cheng2025hawkhierarchicalworkflowframework}, dynamic agent management~\citep{wu2025talkrightspecialistsrouting,zhang2025webpilot}, and end-to-end training~\citep{li2025chainofagentsendtoendagentfoundation,ye2025masgpttrainingllmsbuild}.
However, this escalating architectural complexity invariably introduces significant challenges in maintaining operational robustness and computational efficiency, which we address in this work.

\paragraph{Failure attribution and robustness.}
A significant body of work has emerged to address the challenge of MAS robustness, primarily focusing on post-hoc \textbf{\emph{failure attribution}}~\citep{zhang2025agentcausestaskfailures}.
Systems like Aegis~\citep{song2025aegistaxonomyoptimizationsovercoming} and SHIELDA~\citep{zhou2025shieldastructuredhandlingexceptions} propose taxonomies for failure analysis, while AgenTracer~\citep{zhang2025agentracerinducingfailurellm} and A2P~\citep{west2025abductactpredictscaffolding} introduce methods to better trace the root causes of task failures.
While valuable, these methods are fundamentally reactive, analyzing failures after they have occurred.
In contrast, our \method is designed for \textbf{\emph{proactive, real-time intervention}}, aiming to detect and mitigate high-risk steps \emph{before} they lead to systemic failure.

\paragraph{Efficient Multi-Agent Systems.}
Another stream of research targets the \textbf{efficiency} of MAS, a critical factor largely driven by token consumption. Most approaches focus on \emph{design-time optimization}. Some prune the system's architecture by eliminating agents with AgentDropout~\citep{wang2025agentdropoutdynamicagentelimination} or communication links with SafeSieve~\citep{zhang2025safesieveheuristicsexperienceprogressive}. Others generatively construct efficient prompts~\citep{han2025mapgdmultiagentpromptgradient} or agent topologies from the outset, as seen in MetaAgent~\citep{zhang2025metaagentautomaticallyconstructingmultiagent}, MaAS~\citep{zhang2025agentic-supernet}, and HiVA~\citep{tang2025hivaselforganizedhierarchicalvariable}. A second direction, \emph{context compression}, aims to reduce token count by summarizing or distilling observations~\citep{chen2025smurfs,mou2025ecolangefficienteffectiveagent}. Our work is orthogonal to these methods. Instead of focusing on static design or message content, we introduce \textbf{runtime process control}. \method addresses dynamic inefficiencies \emph{during} execution, a complementary approach that can enhance existing systems.

\section{Preliminary}
\label{sec:preliminary}
In this section, we first establish a formalism for our proposed \underline{S}upervised \underline{M}ulti-\underline{A}gent \underline{S}ystem (SMAS).
We then detail the core components of our framework: the \method's action space and the contextual information it leverages for decision-making.

\subsection{A Formalism for Supervised Multi-Agent Systems}
\label{sec:smas_formalism}
Our work is predicated on the idea that the complex, often chaotic, interactions within a Multi-Agent System (MAS; see Figure \ref{fig:introduction}b) can be actively managed to improve both robustness and efficiency.
To formalize this, we introduce the concept of a {Supervised Multi-Agent System} (SMAS; see Figure \ref{fig:introduction}d).

\begin{definition}[Supervised Multi-Agent System (SMAS)]
    A SMAS is a Multi-Agent System augmented with a meta-level control agent, henceforth referred to as the \textbf{Supervisor}. The Supervisor's objective is to monitor agent interactions in real-time, proactively detecting and mitigating operational risks without altering the core logic of the agents it oversees. In this work, we implement this conceptual Supervisor as a concrete agent named \textbf{\method}.
\end{definition}

The fundamental unit of supervision is the \textbf{interaction}, which occurs when an agent engages with other system components. We categorize interactions into three primary types:
\begin{enumerate}[nosep, leftmargin=12pt]
    \item \textbf{Agent-Agent Interactions:} Communication or delegation between agents.
          In architectures like ReAct~\citep{yao2023reactsynergizingreasoningacting}, where an agent's output becomes another's input, this channel is highly susceptible to the propagation of hallucinated or erroneous information~\citep{shen2025understandinginformationpropagationeffects};
    \item \textbf{Agent-Tool Interactions:}
          The invocation of external tools or APIs.
          This interaction is a primary source of external information, but it is also fraught with risks, including factually incorrect, irrelevant, or outdated data that can corrupt the agent's context~\citep{qian2025smartselfawareagenttool};
    \item \textbf{Agent-Memory Interactions:}
          The retrieval of information from short- or long-term memory stores.
          While crucial for self-evolving systems, memory introduces the hazard of acting upon stale or flawed information from past experiences~\citep{xiong2025memorymanagementimpactsllm}.
\end{enumerate}

\subsection{The \method's Context Window}

To make informed decisions, the \method is provided with a rich, real-time snapshot of the MAS's state, which we formalize as the \emph{context window}.

\begin{definition}[Context Window]
    The standard context window, $\mathcal{W}$, is a tuple of five key elements:
    \[
        \mathcal{W} = (N, Q_g, Q_l, T_l, S) \,,
    \]
    where $N$ is the name of the agent under review, $Q_g$ and $Q_l$ are the global and local tasks, $T_l$ is the \textbf{local trace} of agent $N$'s recent actions and observation summaries, and $S$ is a summary of the agent's latest interaction step.
    For diagnosing system-wide inefficiencies, we augment this to an extended context window $\mathcal{W}_{\text{ext}} = \mathcal{W} \cup \{T_g\}$, where $T_g$ is the \textbf{global trace} of all agent interactions.
\end{definition}

\subsection{The \method's Action Space}
The role of the \method is to diagnose high-risk interactions and execute a targeted intervention (Figure~\ref{fig:methodology}c). We define three primary intervention contexts, $c \in \mathcal{C} = \{c_{\text{error}}, c_{\text{inefficient}}, c_{\text{excessive}} \}$, which activate one of three core supervision strategies:

\begin{itemize}[nosep, leftmargin=12pt,labelindent=0pt]
    \item \textbf{Proactive Error Correction:} Triggered by $c_{\text{error}}$, this strategy aims to diagnose the root cause of an explicit error and provide a direct fix or a verification task to resolve it.
    \item \textbf{Guidance for Inefficiency:} Triggered by $c_{\text{inefficient}}$, this strategy provides pragmatic, course-correcting hints for sub-optimal behaviors, while also critically permitting productive, albeit repetitive, processes to continue via an \textit{approve} action.
    \item \textbf{Adaptive Observation Purification:} Triggered by $c_{\text{excessive}}$, this strategy refines excessively long or noisy observations to improve the signal-to-noise ratio for the agent.
\end{itemize}

These strategies are implemented by selecting an action $a$ from the global action space $\mathcal{A}$. The specific subset of permissible actions, $\mathcal{A}(c)$, is formally defined by the intervention context as follows:
\begin{align*}
    \mathcal{A}(c) =
    \begin{cases}
        \{ \textit{correct\_observation}, \textit{provide\_guidance}, \textit{run\_verification} \} & \text{if } c = c_{\text{error}}       \\
        \{ \textit{approve}, \textit{provide\_guidance} \}                                          & \text{if } c = c_{\text{inefficient}} \\
        \{ \textit{correct\_observation} \}                                                         & \text{if } c = c_{\text{excessive}}
    \end{cases}
\end{align*}
The implementation of each action is detailed in Section~\ref{sec:methodology_action}.

\section{Methodology}
\label{sec:methodology}

\begin{figure}[!t] %
    \centering
    \includegraphics[width=0.9\linewidth]{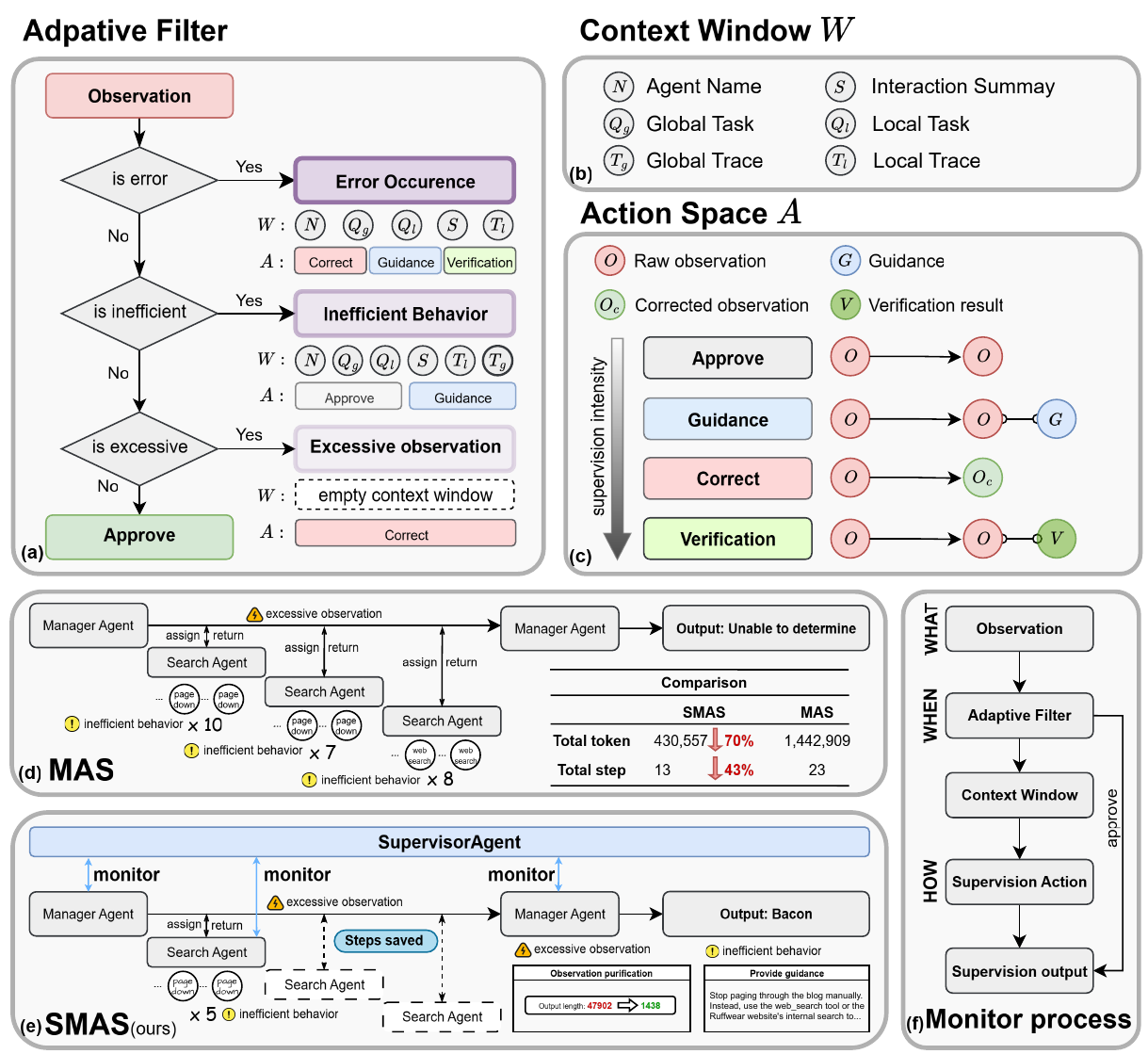}
    \caption{
        \textbf{The architecture and workflow of \method.}
        \textbf{(a)} The LLM-free adaptive filter for identifying high-risk interactions.
        \textbf{(b)} The context window, aggregating goals and traces for situational awareness.
        \textbf{(c)} The spectrum of intervention actions, from simple approval to intensive verification.
        \textbf{(d, e)} Case study on a GAIA task, comparing the baseline MAS (d) with our SMAS (e), which cuts steps by 43\% and token cost by over 70\%.
        \textbf{(f)} The supervise workflow for an interaction, from filtering to a final supervision action.
    }
    \vspace{-1em}
    \label{fig:methodology}
\end{figure}

Building upon the formalism of a Supervised Multi-Agent System (SMAS) introduced in Section~\ref{sec:preliminary}, we now detail the architecture and operational workflow of our \method (illustrated in Figure~\ref{fig:methodology}). Our methodology is structured around three fundamental questions: \textbf{What} to supervise, \textbf{When} to supervise, and \textbf{How} to supervise. We defer the specific implementation details, including all hyperparameters and prompts, to Appendix~\ref{implementation_details} and~\ref{appendix:prompt}.

\subsection{What to Supervise: High-Risk Interaction Points}
The primary targets for our supervision are the three high-risk interaction points defined in our preliminary formalism (Section~\ref{sec:smas_formalism}, see also Figure~\ref{fig:methodology}a): Agent-Agent, Agent-Tool, and Agent-Memory interactions. These points are the primary channels through which errors and inefficiencies are introduced and propagated throughout the system. Our goal is to monitor these specific channels to maintain the operational integrity of the MAS.

\subsection{When to Supervise: The Adaptive Filter}
While a naive approach might monitor every interaction, the associated computational cost is prohibitive and would undermine our goal of improving efficiency.
Therefore, the cornerstone of our framework is a lightweight, LLM-free \textbf{adaptive filter} (see in Figure\ref{fig:methodology}a) designed to trigger supervision only at critical junctures (see case studies in Figures~\ref{fig:methodology}d and~\ref{fig:methodology}e).
This approach ensures that the \method's resources are deployed judiciously, maximizing impact while minimizing overhead.
The filter is designed to be fast and heuristic-based, monitoring the MAS for three pre-defined, high-risk scenarios:
\looseness=-1
\begin{itemize}[nosep, leftmargin=12pt]
    \item \textbf{Error occurrence:}
          The manifestation of an explicit error (e.g., in tool use or code execution) is a critical trigger.
          Unlike current MAS that often pass the full error log into a cluttered context for a subsequent agent to debug, our filter immediately flags these events for a focused, real-time intervention.
          \looseness=-1

    \item \textbf{Inefficient behavior:}
          An agent may enter a loop of sub-optimal or repetitive actions that, while not explicit errors, lead to high token consumption and latency.
          Our filter is designed to detect such patterns, such as an agent repeatedly using the \texttt{page\_down} action instead of a more direct search strategy.
          \looseness=-1

    \item \textbf{Excessive observation length:}
          Interactions with tools can return excessively long and noisy observations (e.g., raw HTML) that inflate costs and distract the agent.
          Our filter identifies such cases for immediate information purification.
          \looseness=-1
\end{itemize}

\subsection{How to Supervise: Memory-Augmented, Multi-Level Intervention}
\label{sec:methodology_action}
Once a high-risk interaction is flagged, \method leverages a rich context window and a spectrum of intervention strategies to deliver a nuanced, effective response.

\paragraph{Memory-augmented context window.} To make an effective decision, a supervisor must possess a more comprehensive understanding of the system's state than any single agent.
This is why \method is conceptualized with its own memory module, not a simple monitor.
As illustrated in Figure~\ref{fig:methodology}b, this is achieved through a dynamic \textbf{context window} $\mathcal{W}$, which aggregates the global task $Q_g$, the agent's local task $Q_l$, interaction summary $S$, and its recent local action trace $T_l$.
Crucially, for diagnosing complex inefficiencies, \method also accesses the \textbf{global trace} $T_g$, granting it a holistic perspective that transcends the limited view of any individual agent.
This elevated viewpoint is what enables it to provide genuinely strategic guidance.

\paragraph{A spectrum of intervention actions.}
With this rich context, \method selects an action from a multi-level action space $\mathcal{A}$, adapting intervention intensity tailored to issue severity (Figure~\ref{fig:methodology}c).
These actions range from a minimal nudge to a comprehensive correction:
\begin{itemize}[nosep, leftmargin=15pt]
    \item \emph{\textbf{approve}}: A minimal intervention that permits a productive, albeit repetitive, agent behavior to continue. Primarily used in the \emph{inefficient} context, its purpose is to avoid disrupting a process that is pragmatically the best path forward from its current state.
    \item \emph{\textbf{provide\_guidance}}: A semi-intrusive action that steers an agent away from a sub-optimal strategy or logical flaw. This action appends a concise, directive hint to the existing observation, correcting the agent's reasoning path without altering the core context data.
    \item \emph{\textbf{correct\_observation}}: A direct and forceful intervention that refines the agent's sensory input. It is the sole action for \emph{excessive observations}, where it purifies the content, and is also used in \emph{error} contexts to fix factually incorrect data.
          This action replaces the original raw observation entirely with a cleaned and corrected version.
    \item \emph{\textbf{run\_verification}}: The deepest intervention, used in complex \emph{error} contexts when internal information is insufficient.
          It invokes a verification sub-agent for external fact-checking or advanced debugging, returning a definitive, verified result.
\end{itemize}

\section{Experiments}
\label{sec:experiments}

\subsection{Experimental Setup}
We empirically validate the effectiveness of \method through a series of extensive experiments.
We begin by outlining our evaluation metrics, datasets, and baselines.
For a more detailed description of the experimental settings, please refer to Appendix \ref{experiment_details}.

\paragraph{Datasets.}
We evaluate our method on a diverse suite of six benchmarks spanning three domains.
Our primary benchmark is the challenging GAIA validation set~\citep{mialon2023gaiabenchmarkgeneralai}, which provides a comprehensive test of an MAS's general problem-solving capabilities.
To demonstrate broader applicability, we use five additional benchmarks: for mathematical reasoning, we use AIME 2024~\citep{HuggingFaceH4_AIME_2024} and a random subset of 600 samples from GSM8k-Hard~\citep{gao2022pal}; for code generation, we use the full HumanEval~\citep{chen2021evaluatinglargelanguagemodels} and MBPP~\citep{austin2021programsynthesislargelanguage} datasets; and for question answering, we use a subset of 800 samples from the DROP~\citep{dua-etal-2019-drop} dataset, following the sampling strategy of prior work~\citep{zhang2025aflowautomatingagenticworkflow}.

\paragraph{Baselines.}
On several benchmarks, we compare \method against a comprehensive set of agentic systems equipped with web-browsing and code execution capabilities.
These baselines fall into two categories: (1) single agent execution methods: including vanilla LLM, Self Consistency CoT (3 answers)~\citep{wang2023selfconsistencyimproveschainthought}, and CodeAgent~\citep{smolagents}; and (2) multi-agent systems, including Smolagent~\citep{smolagents}, OAgents~\citep{zhu2025oagentsempiricalstudybuilding}, MetaAgent~\citep{zhang2025metaagentautomaticallyconstructingmultiagent}, OWL (role playing)~\citep{hu2025owl}, and AWorld~\citep{xie2025profileawaremaneuveringdynamicmultiagent,yu2025aworldorchestratingtrainingrecipe}. 
Detailed descriptions of these baselines are provided in Appendix~\ref{appendix:baselines}.

\paragraph{Implementation details.}
To assess model-agnosticism, we test \method with multiple foundation models.
For the demanding GAIA benchmark, we primarily use GPT-4.1 as the base model for all agents, and evaluate \method when powered by GPT-4.1~\citep{openai_gpt4.1_2025}, Gemini-2.5-pro-0605~\citep{comanici2025gemini25pushingfrontier}, and Qwen3-235B-2507~\citep{qwen3technicalreport}.
For all other benchmarks, we employ the efficient and powerful Qwen3-32B~\citep{qwen3technicalreport} for both the base agents and the \method to assess performance in a more resource-constrained setting.

\paragraph{Testbed selection.}
We selected Smolagent as our primary experimental testbed, which provides a flexible framework upon which we build our agentic systems (SMAS). Critically, Smolagent's capabilities stem primarily from its internal agentic interactions rather than powerful external tools(e.g. web APIs or solvers). This provides an ideal, controlled environment to isolate and evaluate the direct impact of our \method on an agent's core reasoning and communication processes.

\paragraph{Metrics.}
For GAIA and the code generation benchmarks, we report the standard pass@k metric.
For our main baseline, Smolagent, we report pass@1, 2, and 3.
For math reasoning, we report the final solve rate (\%).
For question answering, we report the F1 score for DROP.
In all experiments, we meticulously track and report the total token consumption as a primary measure of efficiency.

\begin{table}[!t]
    \caption{
        \textbf{Overall performance on the GAIA validation set. }
        Our SMAS consistently reduces the average token cost comparing to Smolagent baseline while achieving competitive pass@k success rates.
    }
    \label{tab:multiagent-results}
    \centering
    \resizebox{\textwidth}{!}{
        \begin{tabular}{@{}lcccccccc@{}}
            \toprule
            \multicolumn{1}{c}{\bf Method}                & \multicolumn{1}{c}{\bf Avg. Acc.}              & \multicolumn{1}{c}{\bf Avg. Tokens (K)}        & \multicolumn{1}{c}{\bf L1 Acc.} & \multicolumn{1}{c}{\bf L1 Tokens (K)} & \multicolumn{1}{c}{\bf L2 Acc.} & \multicolumn{1}{c}{\bf L2 Tokens (K)} & \multicolumn{1}{c}{\bf L3 Acc.} & \multicolumn{1}{c}{\bf L3 Tokens (K)} \\
            \midrule
            CodeAgent                                     & 40.00                                          & 120.40                                         & 56.60                           & 92.84                                 & 34.88                           & 131.90                                & 23.08                           & 138.54                                \\
            OWL                                           & 45.40                                          & 111.07                                         & 56.56                           & 67.72                                 & 43.02                           & 110.36                                & 29.16                           & 209.34                                \\
            OAgents                                       & 49.09                                          & 340.50                                         & 66.04                           & 260.27                                & 47.67                           & 358.63                                & 19.23                           & 444.11                                \\
            Smolagent                                     & 50.91                                          & 527.76                                         & 62.26                           & 298.51                                & 53.49                           & 619.59                                & 19.23                           & 691.33                                \\
            AWorld                                        & 60.00                                          & 128.27                                         & 67.92                           & 69.61                                 & 62.79                           & 164.08                                & 34.62                           & 133.65                                \\
            \midrule
            \multicolumn{9}{c}{\bf pass@1}                                                                                                                                                                                                                                                                                                                                                \\
            \midrule
            Smolagent                                     & 50.91                                          & 527.76                                         & 62.26                           & 298.51                                & 53.49                           & 619.59                                & 19.23                           & 691.33                                \\
            \, + \textbf{SMAS (ours)}                     & 50.91                                          & 371.12 {\scriptsize \textcolor{red}{↓29.68\%}} &
            62.26                                         & 258.28 {\scriptsize \textcolor{red}{↓13.48\%}} &
            51.16                                         & 404.96 {\scriptsize \textcolor{red}{↓34.64\%}} &
            26.92 {\scriptsize \textcolor{blue}{↑7.69\%}} & 489.22 {\scriptsize \textcolor{red}{↓29.23\%}}                                                                                                                                                                                                                                                                                \\
            \midrule
            \multicolumn{9}{c}{\bf pass@2}                                                                                                                                                                                                                                                                                                                                                \\
            \midrule
            Smolagent                                     & 58.18                                          & 467.19                                         & 69.81                           & 275.85                                & 59.30                           & 548.02                                & 30.77                           & 589.92                                \\
            \, + \textbf{SMAS (ours)}                     & 58.79 {\scriptsize \textcolor{blue}{↑0.61\%}}  & 389.54 {\scriptsize \textcolor{red}{↓16.62\%}} &
            73.58 {\scriptsize \textcolor{blue}{↑3.77\%}} & 270.07 {\scriptsize \textcolor{red}{↓2.10\%}}  &
            56.98                                         & 420.97 {\scriptsize \textcolor{red}{↓23.18\%}} &
            34.62 {\scriptsize \textcolor{blue}{↑3.85\%}} & 529.20 {\scriptsize \textcolor{red}{↓10.29\%}}                                                                                                                                                                                                                                                                                \\
            \midrule
            \multicolumn{9}{c}{\bf pass@3}                                                                                                                                                                                                                                                                                                                                                \\
            \midrule
            Smolagent                                     & 61.82                                          & 502.40                                         & 71.70                           & 282.14                                & 63.95                           & 605.05                                & 34.62                           & 611.87                                \\
            \, + \textbf{SMAS (ours)}                     & 63.03 {\scriptsize \textcolor{blue}{↑1.21\%}}  & 369.52 {\scriptsize \textcolor{red}{↓26.45\%}} &
            75.47 {\scriptsize \textcolor{blue}{↑3.77\%}} & 276.84 {\scriptsize \textcolor{red}{↓1.88\%}}  &
            62.79                                         & 409.05 {\scriptsize \textcolor{red}{↓32.39\%}} &
            38.46 {\scriptsize \textcolor{blue}{↑3.84\%}} & 427.72 {\scriptsize \textcolor{red}{↓30.10\%}}                                                                                                                                                                                                                                                                                \\
            \bottomrule
        \end{tabular}
    }
\end{table}

\subsection{Results and Analysis}
\paragraph{Significant efficiency gains with competitive accuracy.}
The main experimental results, presented in Table \ref{tab:multiagent-results}, confirm the substantial benefits of \method.
On the GAIA validation set, when integrated with the Smolagent framework, \method achieves an average token reduction of \textbf{29.68\%} at pass@1, while maintaining a statistically equivalent success rate.
Notably, the efficiency gains are even more pronounced on more difficult tasks, with token savings reaching \textbf{32.39\%} on Level 2 and \textbf{30.10\%} on Level 3 tasks at pass@3.

\emph{Across the other five benchmarks, \method generally achieves a Pareto improvement} (see Table \ref{tab:benchmark-generalization}).
In mathematical reasoning, it raises the AIME solve rate by \textbf{6.67\%} while cutting token costs by \textbf{18.92\%}.
In code generation, it maintains competitive accuracy on HumanEval and further reduces token use by \textbf{23.74\%}, likely due to its ability to streamline repetitive debugging cycles.
Occasionally, \method may overcompress long contexts during purification, causing minor accuracy or F1 drops on certain benchmarks.
These results underscore \method's ability to act as a universal efficiency enhancer across diverse problem domains.
\looseness=-1

\begin{table}[!t]
    \caption{
        \textbf{Generalization across diverse benchmarks.}
        \method consistently reduces token costs while maintaining or improving accuracy on tasks spanning mathematical reasoning, code generation, and question answering. All reported gains are relative to the Smolagent baseline.
    }
    \label{tab:benchmark-generalization}
    \centering
    \resizebox{\textwidth}{!}{%
        \begin{tabular}{@{}llccccc@{}}
            \toprule
            \textbf{Method}                            & \textbf{Metrics} & \textbf{GSM-hard}                            & \textbf{AIME}                                 & \textbf{HumanEval}                            & \textbf{MBPP}                                 & \textbf{DROP}                                \\ \midrule
            \multirow{2}{*}{Vanilla}                   & Acc / F1 (\%)    & 67.17                                        & 26.67                                         & 76.82                                         & 80.09                                         & 76.36                                        \\
                                                       & Avg. Tokens (K)  & 0.37                                         & 2.01                                          & 0.28                                          & 0.27                                          & 0.46                                         \\ \cmidrule(l){2-7}
            \multirow{2}{*}{CoT SC (3-shot)}           & Acc / F1 (\%)    & 69.01                                        & 30.00                                         & 77.78                                         & 81.26                                         & 77.72                                        \\
                                                       & Avg. Tokens (K)  & 2.62                                         & 14.26                                         & 1.42                                          & 1.29                                          & 2.73                                         \\ \cmidrule(l){2-7}
            \multirow{2}{*}{OWL}                       & Acc / F1 (\%)    & 72.48                                        & 33.33                                         & 90.74                                         & 79.08                                         & 79.85                                        \\
                                                       & Avg. Tokens (K)  & 15.67                                        & 56.11                                         & 31.87                                         & 54.80                                         & 11.47                                        \\ \cmidrule(l){2-7}
            \multirow{2}{*}{MetaAgent}                 & Acc / F1 (\%)    & 72.14                                        & 26.67                                         & 74.08                                         & 79.86                                         & 78.16                                        \\
                                                       & Avg. Tokens (K)  & 4.35                                         & 6.24                                          & 2.59                                          & 6.39                                          & 1.43                                         \\ \midrule
            \multirow{2}{*}{Smolagent}                 & Acc / F1 (\%)    & 74.33                                        & 30.00                                         & 92.07                                         & 85.68                                         & 81.08                                        \\
                                                       & Avg. Tokens (K)  & 11.59                                        & 59.14                                         & 40.91                                         & 111.07                                        & 12.01                                        \\ \cmidrule(l){2-7}
            \multirow{2}{*}{\, + \textbf{SMAS (ours)}} & Acc / F1 (\%)    & 75.50                                        & 36.67                                         & 92.68                                         & 84.43                                         & 79.80                                        \\
                                                       & Avg. Tokens (K)  & 10.55 {\scriptsize \textcolor{red}{↓8.92\%}} & 47.95 {\scriptsize \textcolor{red}{↓18.92\%}} & 31.19 {\scriptsize \textcolor{red}{↓23.74\%}} & 103.71 {\scriptsize \textcolor{red}{↓6.62\%}} & 11.34 {\scriptsize \textcolor{red}{↓5.60\%}} \\ \bottomrule
        \end{tabular}%
    }
\end{table}

\begin{figure}[!t] %
    \centering
    \includegraphics[width=0.9\linewidth]{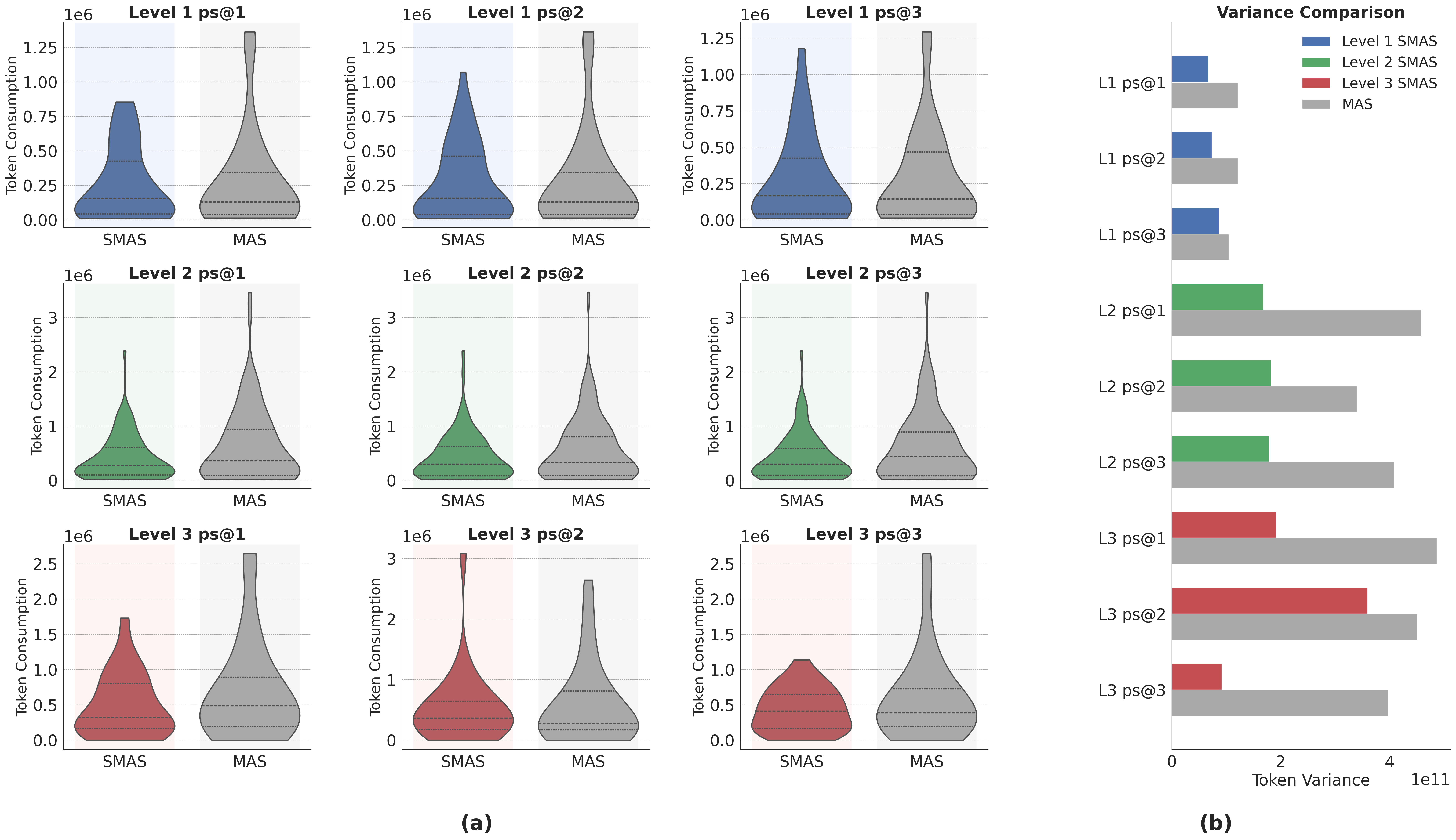}
    \vspace{-0.5em}
    \caption{
        \textbf{\method enhances performance consistency on the GAIA benchmark.}
        \textbf{(a)} Violin plots of token cost distributions, revealing the more compact and predictable performance of our Supervised MAS (SMAS).
        \textbf{(b)} A direct comparison quantifying the substantial reduction in token cost variance achieved by our SMAS across all difficulty levels.
    }
    \label{fig:violin}
\end{figure}

\begin{figure}[!t]
    \centering
    \begin{subfigure}[t]{0.45\textwidth}
        \centering
        \includegraphics[width=\linewidth]{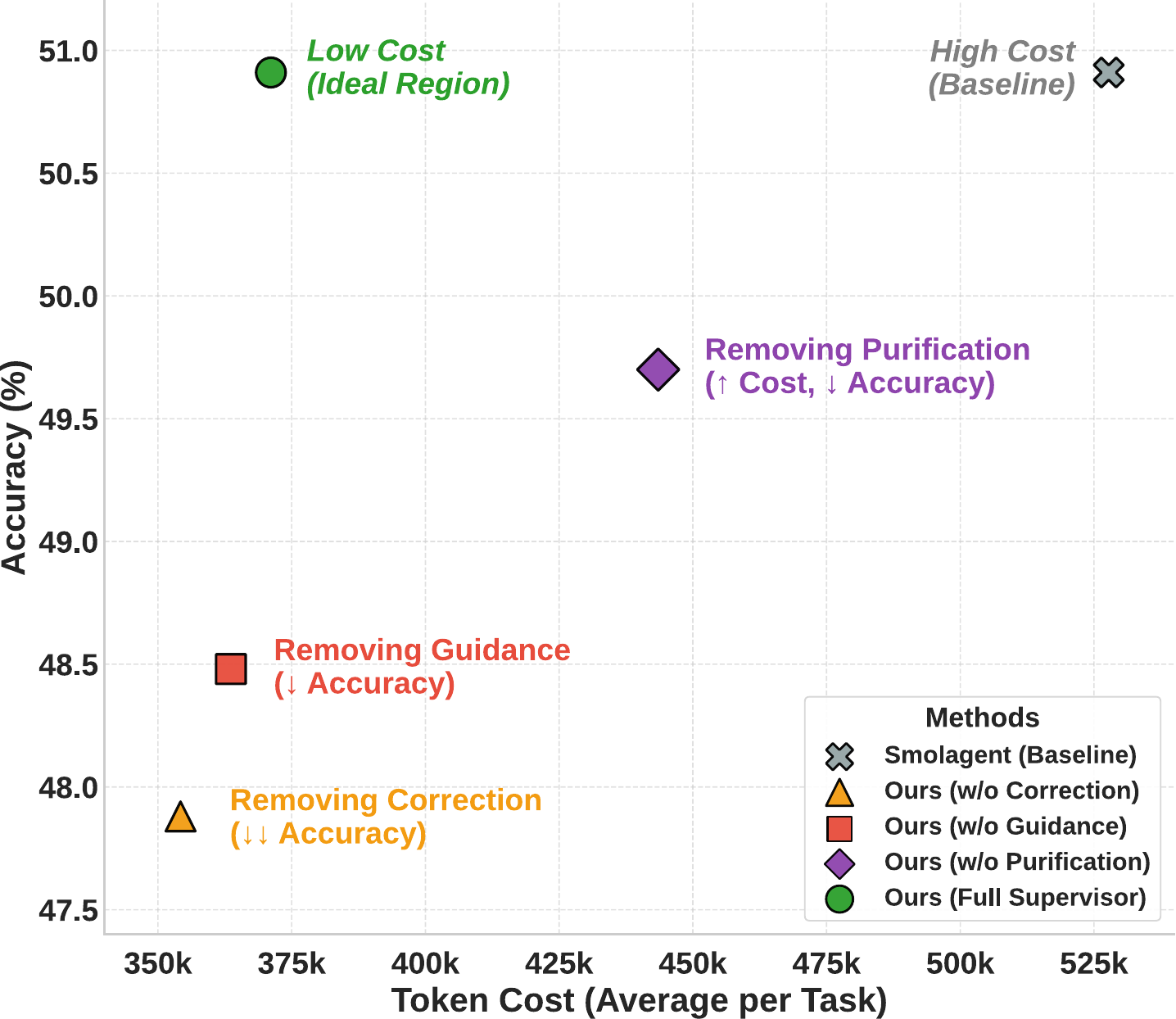}
        \caption{Ablation study on GAIA.}
        \label{fig:ablation}
    \end{subfigure}
    \hfill
    \begin{subfigure}[t]{0.45\textwidth}
        \centering
        \includegraphics[width=\linewidth]{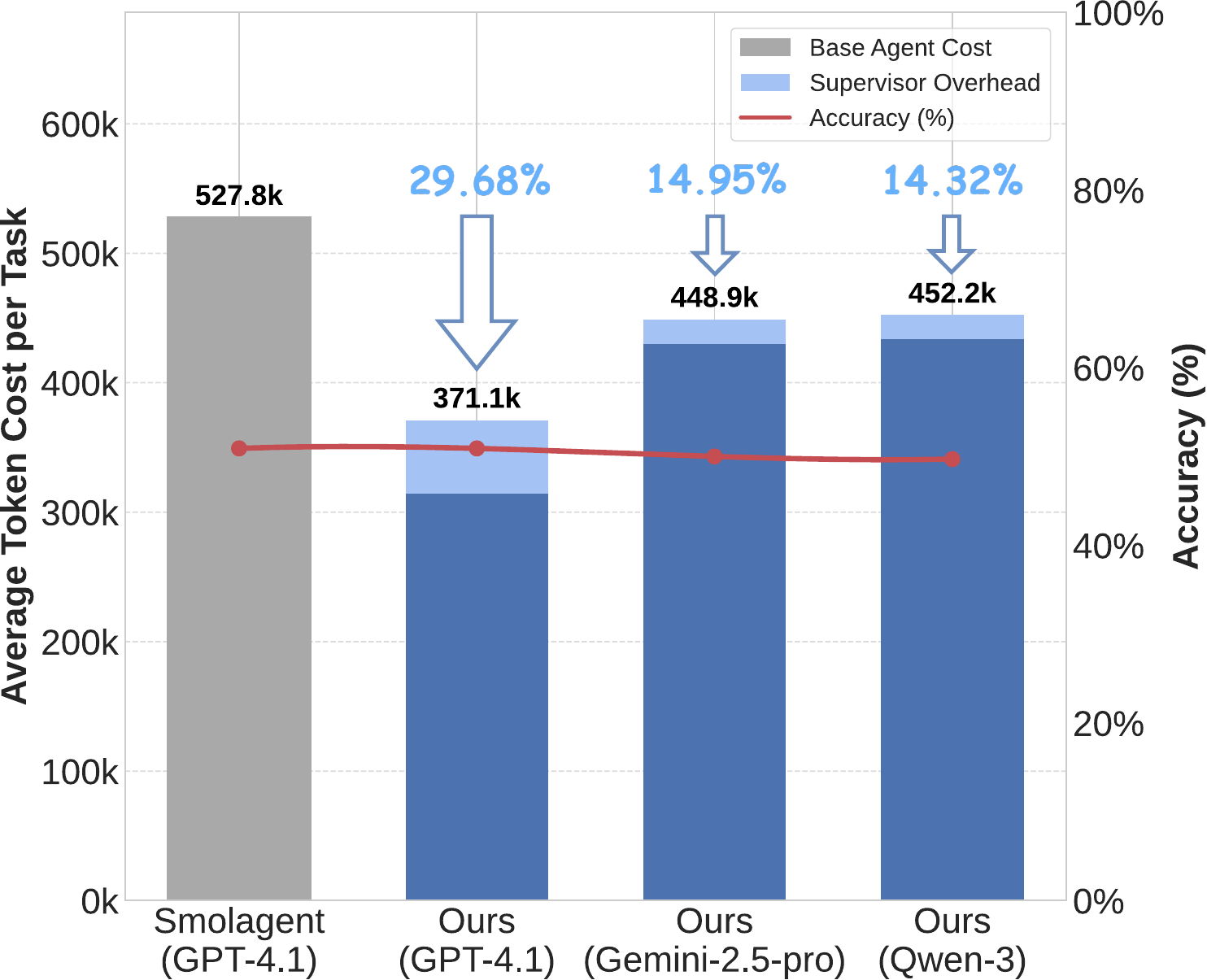}
        \caption{Model Generalization of \method.}
        \label{fig:model-agnostic}
    \end{subfigure}
    \caption{
        \textbf{Ablation study and model generalization of \method.}
        \textbf{(a)}~Ablation study on challenging GAIA tasks, dissecting the distinct contributions of each module to the framework's overall efficiency and robustness.
        \textbf{(b)}~Validation of model-agnosticism, showing that \method consistently delivers token savings across diverse foundation models.
    }
    \label{fig:two-figs}
\end{figure}

\paragraph{Model-Agnostic generalization.}
To demonstrate that the benefits of \method are architectural rather than model-specific, we evaluated it with three different powerful LLMs as its inference engine on GAIA.
As shown in Figure \ref{fig:model-agnostic}, \emph{\method consistently yields significant token savings and maintains robust performance across all models, including GPT-4.1, Gemini-2.5-pro, and Qwen3-235B.}
This validates that our supervision framework is a model-agnostic component that can enhance a wide variety of LLM-powered agent systems.

\paragraph{Improving robustness and performance consistency.}
Beyond average performance, we define robustness as the consistency of an agent's performance.
As illustrated by the violin plots in Figure \ref{fig:violin}, \emph{\method significantly reduces the variance in token consumption per task.}
The distributions for the SMAS are visibly shorter and wider, indicating a more concentrated and predictable performance profile.
The bar chart on the right further quantifies this, showing a marked decrease in token cost variance, especially for the more complex Level 2 and 3 tasks.
This demonstrates that \emph{our method not only makes the MAS more efficient on average but also more reliable and less prone to extreme resource consumption outliers.}

\paragraph{Ablation study.} We conducted an ablation study on the full GAIA validation set to isolate the impact of \method's three core strategies (Table~\ref{tab:ablation_full}, Figure~\ref{fig:ablation}). A comparison of the full framework with \textbf{w/o Correction} (Proactive Error Correction), \textbf{w/o Guidance} (Guidance for Inefficiency), and \textbf{w/o Purification} (Adaptive Observation Purification) reveals distinct roles. \textbf{Purification} is the primary driver of efficiency; disabling it drastically reduces token savings (from 29.68\% to 15.96\%). Conversely, removing \textbf{Correction} or \textbf{Guidance} results in the most significant accuracy drops, confirming their necessity for robustness. This underscores a synergistic design: while Purification minimizes cost, Correction and Guidance ensure task success, justifying their marginal overhead. These benefits are particularly pronounced on high-cost tasks (see Appendix \ref{ablation_study_subset}).

\begin{table}[!t]
    \caption{
        \textbf{Ablation study of \method's components} on the full GAIA validation set.
    }
    \label{tab:ablation_full}
    \centering
    \resizebox{\textwidth}{!}{
        \begin{tabular}{>{\raggedright\arraybackslash}m{4cm} c c c c c}
            \toprule
            \bf Method                   & \bf Avg. Acc. & \bf Avg. Token                                   & \bf Level 1 Avg. Token                           & \bf Level 2 Avg. Token                           & \bf Level 3 Avg. Token                           \\
            \midrule
            Smolagent                    & 50.91         & 527,759                                          & 298,506                                          & 619,591                                          & 691,331                                          \\
            \, + SMAS (w/o Correction)   & 47.88         & 354,226 {\scriptsize \textcolor{blue}{↓32.88\%}} & 221,515 {\scriptsize \textcolor{blue}{↓25.79\%}} & 363,871 {\scriptsize \textcolor{blue}{↓41.27\%}} & 592,852 {\scriptsize \textcolor{blue}{↓14.24\%}} \\
            \, + SMAS (w/o Guidance)     & 48.48         & 363,644 {\scriptsize \textcolor{blue}{↓31.10\%}} & 253,591 {\scriptsize \textcolor{blue}{↓15.05\%}} & 419,913 {\scriptsize \textcolor{blue}{↓32.23\%}} & 401,861 {\scriptsize \textcolor{blue}{↓41.87\%}} \\
            \, + SMAS (w/o Purification) & 49.70         & 443,520 {\scriptsize \textcolor{blue}{↓15.96\%}} & 270,058 {\scriptsize \textcolor{blue}{↓9.53\%}}  & 502,937 {\scriptsize \textcolor{blue}{↓18.83\%}} & 600,582 {\scriptsize \textcolor{blue}{↓13.13\%}} \\
            \, + SMAS                    & 50.91         & 371,119 {\scriptsize \textcolor{blue}{↓29.68\%}} & 258,279 {\scriptsize \textcolor{blue}{↓13.48\%}} & 404,955 {\scriptsize \textcolor{blue}{↓34.64\%}} & 489,222 {\scriptsize \textcolor{blue}{↓29.23\%}} \\
            \bottomrule
        \end{tabular}
    }
\end{table}

\paragraph{MAS-Agnostic generalization.}
To verify MAS-agnostic feature of \method, we integrate \method into two distinct multi-agent system frameworks: AWorld~\citep{xie2025profileawaremaneuveringdynamicmultiagent} and OAgents~\citep{zhu2025oagentsempiricalstudybuilding}, and evaluate their performance on the subset of GAIA benchmark(top-10 most token-intensive tasks per GAIA level).
The results, presented in Table~\ref{tab:mas-agnostic}, indicate that \method consistently enhances the performance of both frameworks, underscoring its versatility and effectiveness across different MAS architectures.

Specifically, integrated with AWorld~\citep{xie2025profileawaremaneuveringdynamicmultiagent}, our SMAS(AWorld) achieved superior average accuracy over both the original AWorld (without Guard) and the Guard-enabled version. Furthermore, SMAS(AWorld) demonstrated substantial token efficiency, saving \textbf{36.54\%} on average versus AWorld (with Guard). With savings reaching \textbf{48.38\%} on Level 3 tasks, it confirms SMAS's ability to enhance tool-intensive MAS.
Applying \method to OAgents~\citep{zhu2025oagentsempiricalstudybuilding} further validated its general applicability, reducing average token consumption by \textbf{39.36\%} while maintaining competitive accuracy. Interestingly, the largest token reduction (\textbf{50.19\%}) occurred on Level 1 tasks, exceeding the savings on Level 3 (\textbf{40.63\%}). While this might reflect OAgents' inherent proficiency on harder tasks, the significant overall savings underscore \method's broad utility.

\paragraph{Overhead analysis.}
Crucially, all efficiency gains reported in this work represent \textbf{net savings}, fully accounting for the cost of \method. As detailed in Table \ref{tab:super_overhead}, the supervisor itself incurs a modest overhead, averaging only \textbf{15.45\%} of total token usage, which validates its lightweight design. Regarding latency, the supervisory interventions introduce an average increase of less than one minute and a half per task (Table \ref{tab:ablation_latency}). We consider this temporal cost a justifiable trade-off given the substantial economic savings achieved in complex multi-agent workflows. A comprehensive overhead analysis is provided in Appendix \ref{overhead_analysis}.

\begin{table}[!t]
    \caption{
        \textbf{Cross-framework performance of \method.}
        Evaluated on GAIA subset (top-10 most token-intensive tasks per level).
    }
    \label{tab:mas-agnostic}
    \centering
    \resizebox{\textwidth}{!}{
        \begin{tabular}{>{\raggedright\arraybackslash}m{4cm} c c c c c}
            \toprule
            \bf Method                  & \bf Avg. Acc.                                & \bf Avg. Token                                   & \bf Level 1 Avg. Token                           & \bf Level 2 Avg. Token                           & \bf Level 3 Avg. Token                           \\
            \midrule
            Smolagent                   & 40.00                                        & 1,446,526                                        & 933,013                                          & 2,037,437                                        & 1,369,131                                        \\
            \textbf{\, + SMAS}          & 46.67 {\scriptsize \textcolor{red}{↑6.67\%}} & 721,332 {\scriptsize \textcolor{blue}{↓50.13\%}} & 522,364 {\scriptsize \textcolor{blue}{↓44.01\%}} & 960,694 {\scriptsize \textcolor{blue}{↓52.85\%}} & 680,939 {\scriptsize \textcolor{blue}{↓50.26\%}} \\ \midrule
            AWorld (without Guard)      & 23.33                                        & 155,239                                          & 50,851                                           & 217,332                                          & 166,500                                          \\
            AWorld (with Guard)         & 30.00                                        & 353,738                                          & 135,413                                          & 463,083                                          & 376,878                                          \\
            \textbf{AWorld (with SMAS)} & 36.67 {\scriptsize \textcolor{red}{↑6.67\%}} & 224,480 {\scriptsize \textcolor{blue}{↓36.54\%}} & 90,569 {\scriptsize \textcolor{blue}{↓33.12\%}}  & 355,051 {\scriptsize \textcolor{blue}{↓23.33\%}} & 194,561 {\scriptsize \textcolor{blue}{↓48.38\%}} \\ \midrule
            OAgents                     & 46.67                                        & 530,939                                          & 430,852                                          & 359,511                                          & 802,454                                          \\
            \textbf{\, + SMAS}          & 46.67                                        & 321,957 {\scriptsize \textcolor{blue}{↓39.36\%}} & 214,604 {\scriptsize \textcolor{blue}{↓50.19\%}} & 274,875 {\scriptsize \textcolor{blue}{↓23.54\%}} & 476,393 {\scriptsize \textcolor{blue}{↓40.63\%}} \\
            \bottomrule
        \end{tabular}
    }
\end{table}

\section{Discussion and Conclusion}

\paragraph{Supervisor as a foundational MAS component.}

Our work positions \method as a foundational component for future Multi-Agent Systems, akin to established modules like memory banks and tool-usage frameworks. By providing real-time, adaptive supervision, \method alleviates critical challenges of robustness and efficiency that are pervasive across diverse MAS architectures. Its modular design allows for seamless integration with existing systems, enhancing their performance without necessitating fundamental changes to their core logic. This underscores the potential of supervisory agents as universal enhancers of MASs, capable of elevating both reliability and cost-effectiveness across a wide range of applications.

\paragraph{Comparison with related supervisory agents.}
A related concept is the Guard agent in AWorld~\citep{xie2025profileawaremaneuveringdynamicmultiagent}, which is invoked at key steps primarily for factual verification to enhance task accuracy. While valuable, its scope differs significantly from our method. \method{} adopts a broader objective of improving overall system efficiency and robustness through continuous (albeit adaptively filtered) monitoring and a wider range of interventions, including error correction, inefficiency guidance, and observation purification, complementing the Guard's focus on accuracy.

\paragraph{Broader insights.}

Our work also yields critical insights for the broader field. First, we discovered that seemingly “noisy” information, such as HTML structure and truncation cues, serves as a vital signal for ReAct-style agents. The overly aggressive purification can paradoxically harm performance. This highlights a fundamental trade-off between information density and the preservation of environmental texture.
Second, our focus on token cost underscores the need for a more holistic efficiency evaluation for MAS. A comprehensive analysis must also account for the frequency and complexity of external tool API calls, which offload significant burdens from the MAS.
This very trade-off informed our choice of Smolagent as a primary testbed - its reliance on internal agentic reasoning, rather than powerful external tools, provided a controlled environment to isolate and evaluate our \method's impact on the interaction process itself.

\paragraph{Future directions.}
These insights inform several promising avenues for future work. First, moving beyond heuristic rules, exploring a learning-based adaptive filter could enable more precise, dynamic control over supervisor invocations. This aligns with the broader goal of developing a self-evolving, memory-augmented version of \method{}. Second, further research should focus on mitigating the latency overhead introduced by supervisory calls to enhance real-time applicability, alongside creating sophisticated purification techniques that address the ``noise-as-signal'' trade-off. Finally, developing a universal resource consumption metric for MAS remains a critical open challenge. Ultimately, we posit that incorporating such real-time, meta-level supervision is a foundational component for building the next generation of truly scalable and reliable MAS.

\paragraph{Conclusion.}

In this work, we introduced \textbf{\method}, a lightweight and non-intrusive meta-agent framework that enhances the robustness and efficiency of Multi-Agent Systems. Through real-time, adaptive supervision, \method{} mitigates common failure modes and reduces computational overhead using three core strategies: proactive error correction, pragmatic inefficiency guidance, and adaptive observation purification. Our extensive experiments demonstrate a significant Pareto improvement. On the challenging GAIA benchmark, \method{} reduces token consumption by an average of 29.68\% while maintaining competitive task success rates, a crucial step towards building more practical and scalable agentic systems.

\newpage
\section*{Acknowledgements}
This work was supported in part by the National Key Research and Development Program of China (2024YFF0907803), Research Fund for International Scientists of National Natural Science Foundation of China (72350710798), National Natural Science Foundation of China (NSFC) under No.\ 62576285, 62276230, Research Center for Industries of the Future (RCIF) at Westlake University, and Westlake Education Foundation.

\section*{Ethics Statement}
Our work aims to improve the reliability and efficiency of Multi-Agent Systems, a crucial step for developing practical and beneficial autonomous technologies. We believe that by introducing a mechanism for real-time supervision, our framework provides a paradigm not only for performance optimization but also for enhancing the safety and predictability of future agentic systems. Our research was conducted on publicly available benchmarks, did not involve private user data, and adheres to the ICLR Code of Ethics.

\section*{Reproducibility Statement}
We are committed to ensuring our work is reproducible. The core architecture and logic of \method{} are detailed in Section~\ref{sec:methodology}, with theoretical formalisms in Section~\ref{sec:preliminary}. For direct replication, we provide all implementation details and final prompts in Appendix~\ref{implementation_details},~\ref{appendix:prompt}, and our code is available at \url{https://github.com/LINs-lab/SupervisorAgent}. The datasets and metrics used in our extensive experiments (Section~\ref{sec:experiments}) are all based on publicly available benchmarks, allowing for direct comparison and validation of our results.

\bibliography{resources/reference}

@misc{liu2025advanceschallengesfoundationagents,
      title={Advances and Challenges in Foundation Agents: From Brain-Inspired Intelligence to Evolutionary, Collaborative, and Safe Systems}, 
      author={Bang Liu and Xinfeng Li and Jiayi Zhang and Jinlin Wang and Tanjin He and Sirui Hong and Hongzhang Liu and Shaokun Zhang and Kaitao Song and Kunlun Zhu and Yuheng Cheng and Suyuchen Wang and Xiaoqiang Wang and Yuyu Luo and Haibo Jin and Peiyan Zhang and Ollie Liu and Jiaqi Chen and Huan Zhang and Zhaoyang Yu and Haochen Shi and Boyan Li and Dekun Wu and Fengwei Teng and Xiaojun Jia and Jiawei Xu and Jinyu Xiang and Yizhang Lin and Tianming Liu and Tongliang Liu and Yu Su and Huan Sun and Glen Berseth and Jianyun Nie and Ian Foster and Logan Ward and Qingyun Wu and Yu Gu and Mingchen Zhuge and Xinbing Liang and Xiangru Tang and Haohan Wang and Jiaxuan You and Chi Wang and Jian Pei and Qiang Yang and Xiaoliang Qi and Chenglin Wu},
      year={2025},
      eprint={2504.01990},
      archivePrefix={arXiv},
      primaryClass={cs.AI},
      url={https://arxiv.org/abs/2504.01990}, 
}

@misc{gao2025surveyselfevolvingagentspath,
      title={A Survey of Self-Evolving Agents: On Path to Artificial Super Intelligence}, 
      author={{Huan-ang} Gao and Jiayi Geng and Wenyue Hua and Mengkang Hu and Xinzhe Juan and Hongzhang Liu and Shilong Liu and Jiahao Qiu and Xuan Qi and Yiran Wu and Hongru Wang and Han Xiao and Yuhang Zhou and Shaokun Zhang and Jiayi Zhang and Jinyu Xiang and Yixiong Fang and Qiwen Zhao and Dongrui Liu and Qihan Ren and Cheng Qian and Zhenghailong Wang and Minda Hu and Huazheng Wang and Qingyun Wu and Heng Ji and Mengdi Wang},
      year={2025},
      eprint={2507.21046},
      archivePrefix={arXiv},
      primaryClass={cs.AI},
      url={https://arxiv.org/abs/2507.21046}, 
}

@misc{shang2025rstar2agentagenticreasoningtechnical,
      title={rStar2-Agent: Agentic Reasoning Technical Report}, 
      author={Ning Shang and Yifei Liu and Yi Zhu and Li Lyna Zhang and Weijiang Xu and Xinyu Guan and Buze Zhang and Bingcheng Dong and Xudong Zhou and Bowen Zhang and Ying Xin and Ziming Miao and Scarlett Li and Fan Yang and Mao Yang},
      year={2025},
      eprint={2508.20722},
      archivePrefix={arXiv},
      primaryClass={cs.CL},
      url={https://arxiv.org/abs/2508.20722}, 
}

@misc{lu2025requirementsdevelopmentformalizationreliable,
      title={Requirements Development and Formalization for Reliable Code Generation: A Multi-Agent Vision}, 
      author={Xu Lu and Weisong Sun and Yiran Zhang and Ming Hu and Cong Tian and Zhi Jin and Yang Liu},
      year={2025},
      eprint={2508.18675},
      archivePrefix={arXiv},
      primaryClass={cs.SE},
      url={https://arxiv.org/abs/2508.18675}, 
}

@misc{luo2025entitylinkingagentquestion,
      title={An Entity Linking Agent for Question Answering}, 
      author={Yajie Luo and Yihong Wu and Muzhi Li and Fengran Mo and Jia Ao Sun and Xinyu Wang and Liheng Ma and Yingxue Zhang and Jian-Yun Nie},
      year={2025},
      eprint={2508.03865},
      archivePrefix={arXiv},
      primaryClass={cs.CL},
      url={https://arxiv.org/abs/2508.03865}, 
}

@misc{shi2025mobileguirladvancingmobilegui,
      title={MobileGUI-RL: Advancing Mobile GUI Agent through Reinforcement Learning in Online Environment}, 
      author={Yucheng Shi and Wenhao Yu and Zaitang Li and Yonglin Wang and Hongming Zhang and Ninghao Liu and Haitao Mi and Dong Yu},
      year={2025},
      eprint={2507.05720},
      archivePrefix={arXiv},
      primaryClass={cs.LG},
      url={https://arxiv.org/abs/2507.05720}, 
}

@misc{liu2025infiguir1advancingmultimodalgui,
      title={InfiGUI-R1: Advancing Multimodal GUI Agents from Reactive Actors to Deliberative Reasoners}, 
      author={Yuhang Liu and Pengxiang Li and Congkai Xie and Xavier Hu and Xiaotian Han and Shengyu Zhang and Hongxia Yang and Fei Wu},
      year={2025},
      eprint={2504.14239},
      archivePrefix={arXiv},
      primaryClass={cs.AI},
      url={https://arxiv.org/abs/2504.14239}, 
}

@misc{li2025assemblecrewautomaticmultiagent,
      title={Assemble Your Crew: Automatic Multi-agent Communication Topology Design via Autoregressive Graph Generation}, 
      author={Shiyuan Li and Yixin Liu and Qingsong Wen and Chengqi Zhang and Shirui Pan},
      year={2025},
      eprint={2507.18224},
      archivePrefix={arXiv},
      primaryClass={cs.MA},
      url={https://arxiv.org/abs/2507.18224}, 
}

@misc{wu2025dissectingadversarialrobustnessmultimodal,
      title={Dissecting Adversarial Robustness of Multimodal LM Agents}, 
      author={Chen Henry Wu and Rishi Shah and Jing Yu Koh and Ruslan Salakhutdinov and Daniel Fried and Aditi Raghunathan},
      year={2025},
      eprint={2406.12814},
      archivePrefix={arXiv},
      primaryClass={cs.LG},
      url={https://arxiv.org/abs/2406.12814}, 
}

@inproceedings{huang2025competing,
  author    = {Jen{-}tse Huang and
               Eric John Li and
               Man Ho Lam and
               Tian Liang and
               Wenxuan Wang and
               Youliang Yuan and
               Wenxiang Jiao and
               Xing Wang and
               Zhaopeng Tu and
               Michael R. Lyu},
  title     = {Competing Large Language Models in Multi-Agent Gaming Environments},
  booktitle = {Proceedings of the Thirteenth International Conference on Learning Representations (ICLR)},
  year      = {2025}
}

@misc{li2025chainofagentsendtoendagentfoundation,
      title={Chain-of-Agents: End-to-End Agent Foundation Models via Multi-Agent Distillation and Agentic RL}, 
      author={Weizhen Li and Jianbo Lin and Zhuosong Jiang and Jingyi Cao and Xinpeng Liu and Jiayu Zhang and Zhenqiang Huang and Qianben Chen and Weichen Sun and Qiexiang Wang and Hongxuan Lu and Tianrui Qin and Chenghao Zhu and Yi Yao and Shuying Fan and Xiaowan Li and Tiannan Wang and Pai Liu and King Zhu and He Zhu and Dingfeng Shi and Piaohong Wang and Yeyi Guan and Xiangru Tang and Minghao Liu and Yuchen Eleanor Jiang and Jian Yang and Jiaheng Liu and Ge Zhang and Wangchunshu Zhou},
      year={2025},
      eprint={2508.13167},
      archivePrefix={arXiv},
      primaryClass={cs.AI},
      url={https://arxiv.org/abs/2508.13167}, 
}

@article{farquhar_detecting_2024,
  title   = {Detecting hallucinations in large language models using semantic entropy},
  author  = {Farquhar, Sebastian and Kossen, Jannik and Kuhn, Lorenz and Gal, Yarin},
  journal = {Nature},
  volume  = {630},
  number  = {8017},
  pages   = {625--630},
  year    = {2024},
  month   = {jun},
  issn    = {1476-4687},
  doi     = {10.1038/s41586-024-07421-0},
  url     = {https://www.nature.com/articles/s41586-024-07421-0}
}

@misc{kalai2025languagemodelshallucinate,
      title={Why Language Models Hallucinate}, 
      author={Adam Tauman Kalai and Ofir Nachum and Santosh S. Vempala and Edwin Zhang},
      year={2025},
      eprint={2509.04664},
      archivePrefix={arXiv},
      primaryClass={cs.CL},
      url={https://arxiv.org/abs/2509.04664}, 
}

@misc{dong2025practicalmemoryinjectionattack,
      title={A Practical Memory Injection Attack against LLM Agents}, 
      author={Shen Dong and Shaochen Xu and Pengfei He and Yige Li and Jiliang Tang and Tianming Liu and Hui Liu and Zhen Xiang},
      year={2025},
      eprint={2503.03704},
      archivePrefix={arXiv},
      primaryClass={cs.LG},
      url={https://arxiv.org/abs/2503.03704}, 
}

@misc{chen2024agentpoisonredteamingllmagents,
      title={AgentPoison: Red-teaming LLM Agents via Poisoning Memory or Knowledge Bases}, 
      author={Zhaorun Chen and Zhen Xiang and Chaowei Xiao and Dawn Song and Bo Li},
      year={2024},
      eprint={2407.12784},
      archivePrefix={arXiv},
      primaryClass={cs.LG},
      url={https://arxiv.org/abs/2407.12784}, 
}

@misc{shen2025understandinginformationpropagationeffects,
      title={Understanding the Information Propagation Effects of Communication Topologies in LLM-based Multi-Agent Systems}, 
      author={Xu Shen and Yixin Liu and Yiwei Dai and Yili Wang and Rui Miao and Yue Tan and Shirui Pan and Xin Wang},
      year={2025},
      eprint={2505.23352},
      archivePrefix={arXiv},
      primaryClass={cs.MA},
      url={https://arxiv.org/abs/2505.23352}, 
}

@misc{tian2025outlookopportunitieschallengesmultiagent,
      title={An Outlook on the Opportunities and Challenges of Multi-Agent AI Systems}, 
      author={Fangqiao Tian and An Luo and Jin Du and Xun Xian and Robert Specht and Ganghua Wang and Xuan Bi and Jiawei Zhou and Ashish Kundu and Jayanth Srinivasa and Charles Fleming and Rui Zhang and Zirui Liu and Mingyi Hong and Jie Ding},
      year={2025},
      eprint={2505.18397},
      archivePrefix={arXiv},
      primaryClass={cs.MA},
      url={https://arxiv.org/abs/2505.18397}, 
}

@inproceedings{hosseini-etal-2025-efficient,
    title = "Efficient Solutions For An Intriguing Failure of {LLM}s: Long Context Window Does Not Mean {LLM}s Can Analyze Long Sequences Flawlessly",
    author = "Hosseini, Peyman  and
      Castro, Ignacio  and
      Ghinassi, Iacopo  and
      Purver, Matthew",
    editor = "Rambow, Owen  and
      Wanner, Leo  and
      Apidianaki, Marianna  and
      Al-Khalifa, Hend  and
      Eugenio, Barbara Di  and
      Schockaert, Steven",
    booktitle = "Proceedings of the 31st International Conference on Computational Linguistics",
    month = jan,
    year = "2025",
    address = "Abu Dhabi, UAE",
    publisher = "Association for Computational Linguistics",
    url = "https://aclanthology.org/2025.coling-main.128/",
    pages = "1880--1891"
}

@misc{wang2025efficientagentsbuildingeffective,
      title={Efficient Agents: Building Effective Agents While Reducing Cost}, 
      author={Ningning Wang and Xavier Hu and Pai Liu and He Zhu and Yue Hou and Heyuan Huang and Shengyu Zhang and Jian Yang and Jiaheng Liu and Ge Zhang and Changwang Zhang and Jun Wang and Yuchen Eleanor Jiang and Wangchunshu Zhou},
      year={2025},
      eprint={2508.02694},
      archivePrefix={arXiv},
      primaryClass={cs.AI},
      url={https://arxiv.org/abs/2508.02694}, 
}

@misc{cemri2025multiagentllmsystemsfail,
      title={Why Do Multi-Agent LLM Systems Fail?}, 
      author={Mert Cemri and Melissa Z. Pan and Shuyi Yang and Lakshya A. Agrawal and Bhavya Chopra and Rishabh Tiwari and Kurt Keutzer and Aditya Parameswaran and Dan Klein and Kannan Ramchandran and Matei Zaharia and Joseph E. Gonzalez and Ion Stoica},
      year={2025},
      eprint={2503.13657},
      archivePrefix={arXiv},
      primaryClass={cs.AI},
      url={https://arxiv.org/abs/2503.13657}, 
}

@misc{mialon2023gaiabenchmarkgeneralai,
      title={GAIA: a benchmark for General AI Assistants}, 
      author={Grégoire Mialon and Clémentine Fourrier and Craig Swift and Thomas Wolf and Yann LeCun and Thomas Scialom},
      year={2023},
      eprint={2311.12983},
      archivePrefix={arXiv},
      primaryClass={cs.CL},
      url={https://arxiv.org/abs/2311.12983}, 
}

@Misc{smolagents,
  title =        {`smolagents`: a smol library to build great agentic systems.},
  author =       {Aymeric Roucher and Albert Villanova del Moral and Thomas Wolf and Leandro von Werra and Erik Kaunismäki},
  howpublished = {\url{https://github.com/huggingface/smolagents}},
  year =         {2025}
}

@inproceedings{zhang2025webpilot,
  title={Webpilot: A versatile and autonomous multi-agent system for web task execution with strategic exploration},
  author={Zhang, Yao and Ma, Zijian and Ma, Yunpu and Han, Zhen and Wu, Yu and Tresp, Volker},
  booktitle={Proceedings of the AAAI Conference on Artificial Intelligence},
  volume={39},
  number={22},
  pages={23378--23386},
  year={2025}
}

@misc{cheng2025hawkhierarchicalworkflowframework,
      title={HAWK: A Hierarchical Workflow Framework for Multi-Agent Collaboration}, 
      author={Yuyang Cheng and Yumiao Xu and Chaojia Yu and Yong Zhao},
      year={2025},
      eprint={2507.04067},
      archivePrefix={arXiv},
      primaryClass={cs.AI},
      url={https://arxiv.org/abs/2507.04067}, 
}

@misc{tran2025multiagentcollaborationmechanismssurvey,
      title={Multi-Agent Collaboration Mechanisms: A Survey of LLMs}, 
      author={Khanh-Tung Tran and Dung Dao and Minh-Duong Nguyen and Quoc-Viet Pham and Barry O'Sullivan and Hoang D. Nguyen},
      year={2025},
      eprint={2501.06322},
      archivePrefix={arXiv},
      primaryClass={cs.AI},
      url={https://arxiv.org/abs/2501.06322}, 
}

@article{he2025llm,
  title={LLM-Based Multi-Agent Systems for Software Engineering: Literature Review, Vision, and the Road Ahead},
  author={He, Junda and Treude, Christoph and Lo, David},
  journal={ACM Transactions on Software Engineering and Methodology},
  volume={34},
  number={5},
  pages={1--30},
  year={2025},
  publisher={ACM New York, NY}
}

@misc{wu2025talkrightspecialistsrouting,
      title={Talk to Right Specialists: Routing and Planning in Multi-agent System for Question Answering}, 
      author={Feijie Wu and Zitao Li and Fei Wei and Yaliang Li and Bolin Ding and Jing Gao},
      year={2025},
      eprint={2501.07813},
      archivePrefix={arXiv},
      primaryClass={cs.MA},
      url={https://arxiv.org/abs/2501.07813}, 
}

@misc{ye2025masgpttrainingllmsbuild,
      title={MAS-GPT: Training LLMs to Build LLM-based Multi-Agent Systems}, 
      author={Rui Ye and Shuo Tang and Rui Ge and Yaxin Du and Zhenfei Yin and Siheng Chen and Jing Shao},
      year={2025},
      eprint={2503.03686},
      archivePrefix={arXiv},
      primaryClass={cs.CL},
      url={https://arxiv.org/abs/2503.03686}, 
}

@misc{zhu2025oagentsempiricalstudybuilding,
      title={OAgents: An Empirical Study of Building Effective Agents}, 
      author={He Zhu and Tianrui Qin and King Zhu and Heyuan Huang and Yeyi Guan and Jinxiang Xia and Yi Yao and Hanhao Li and Ningning Wang and Pai Liu and Tianhao Peng and Xin Gui and Xiaowan Li and Yuhui Liu and Yuchen Eleanor Jiang and Jun Wang and Changwang Zhang and Xiangru Tang and Ge Zhang and Jian Yang and Minghao Liu and Xitong Gao and Jiaheng Liu and Wangchunshu Zhou},
      year={2025},
      eprint={2506.15741},
      archivePrefix={arXiv},
      primaryClass={cs.AI},
      url={https://arxiv.org/abs/2506.15741}, 
}

@misc{zhang2025agentorchestrahierarchicalmultiagentframework,
      title={AgentOrchestra: A Hierarchical Multi-Agent Framework for General-Purpose Task Solving}, 
      author={Wentao Zhang and Liang Zeng and Yuzhen Xiao and Yongcong Li and Ce Cui and Yilei Zhao and Rui Hu and Yang Liu and Yahui Zhou and Bo An},
      year={2025},
      eprint={2506.12508},
      archivePrefix={arXiv},
      primaryClass={cs.AI},
      url={https://arxiv.org/abs/2506.12508}, 
}

@misc{shi2025aimefullyautonomousmultiagentframework,
      title={Aime: Towards Fully-Autonomous Multi-Agent Framework}, 
      author={Yexuan Shi and Mingyu Wang and Yunxiang Cao and Hongjie Lai and Junjian Lan and Xin Han and Yu Wang and Jie Geng and Zhenan Li and Zihao Xia and Xiang Chen and Chen Li and Jian Xu and Wenbo Duan and Yuanshuo Zhu},
      year={2025},
      eprint={2507.11988},
      archivePrefix={arXiv},
      primaryClass={cs.AI},
      url={https://arxiv.org/abs/2507.11988}, 
}

@misc{tongyidr,
  author={Tongyi DeepResearch Team},
  title={Tongyi-DeepResearch},
  year={2025},
  howpublished={\url{https://github.com/Alibaba-NLP/DeepResearch}}
}

@misc{zhang2025agentcausestaskfailures,
      title={Which Agent Causes Task Failures and When? On Automated Failure Attribution of LLM Multi-Agent Systems}, 
      author={Shaokun Zhang and Ming Yin and Jieyu Zhang and Jiale Liu and Zhiguang Han and Jingyang Zhang and Beibin Li and Chi Wang and Huazheng Wang and Yiran Chen and Qingyun Wu},
      year={2025},
      eprint={2505.00212},
      archivePrefix={arXiv},
      primaryClass={cs.MA},
      url={https://arxiv.org/abs/2505.00212}, 
}

@misc{song2025aegistaxonomyoptimizationsovercoming,
      title={Aegis: Taxonomy and Optimizations for Overcoming Agent-Environment Failures in LLM Agents}, 
      author={Kevin Song and Anand Jayarajan and Yaoyao Ding and Qidong Su and Zhanda Zhu and Sihang Liu and Gennady Pekhimenko},
      year={2025},
      eprint={2508.19504},
      archivePrefix={arXiv},
      primaryClass={cs.MA},
      url={https://arxiv.org/abs/2508.19504}, 
}

@misc{zhang2025agentracerinducingfailurellm,
      title={AgenTracer: Who Is Inducing Failure in the LLM Agentic Systems?}, 
      author={Guibin Zhang and Junhao Wang and Junjie Chen and Wangchunshu Zhou and Kun Wang and Shuicheng Yan},
      year={2025},
      eprint={2509.03312},
      archivePrefix={arXiv},
      primaryClass={cs.CL},
      url={https://arxiv.org/abs/2509.03312}, 
}

@misc{zhou2025shieldastructuredhandlingexceptions,
      title={SHIELDA: Structured Handling of Exceptions in LLM-Driven Agentic Workflows}, 
      author={Jingwen Zhou and Jieshan Chen and Qinghua Lu and Dehai Zhao and Liming Zhu},
      year={2025},
      eprint={2508.07935},
      archivePrefix={arXiv},
      primaryClass={cs.SE},
      url={https://arxiv.org/abs/2508.07935}, 
}

@misc{west2025abductactpredictscaffolding,
      title={Abduct, Act, Predict: Scaffolding Causal Inference for Automated Failure Attribution in Multi-Agent Systems}, 
      author={Alva West and Yixuan Weng and Minjun Zhu and Zhen Lin and Yue Zhang},
      year={2025},
      eprint={2509.10401},
      archivePrefix={arXiv},
      primaryClass={cs.AI},
      url={https://arxiv.org/abs/2509.10401}, 
}

@misc{wang2025agentdropoutdynamicagentelimination,
      title={AgentDropout: Dynamic Agent Elimination for Token-Efficient and High-Performance LLM-Based Multi-Agent Collaboration}, 
      author={Zhexuan Wang and Yutong Wang and Xuebo Liu and Liang Ding and Miao Zhang and Jie Liu and Min Zhang},
      year={2025},
      eprint={2503.18891},
      archivePrefix={arXiv},
      primaryClass={cs.CL},
      url={https://arxiv.org/abs/2503.18891}, 
}

@misc{zhang2025safesieveheuristicsexperienceprogressive,
      title={SafeSieve: From Heuristics to Experience in Progressive Pruning for LLM-based Multi-Agent Communication}, 
      author={Ruijia Zhang and Xinyan Zhao and Ruixiang Wang and Sigen Chen and Guibin Zhang and An Zhang and Kun Wang and Qingsong Wen},
      year={2025},
      eprint={2508.11733},
      archivePrefix={arXiv},
      primaryClass={cs.MA},
      url={https://arxiv.org/abs/2508.11733}, 
}

@misc{zhang2025metaagentautomaticallyconstructingmultiagent,
      title={MetaAgent: Automatically Constructing Multi-Agent Systems Based on Finite State Machines}, 
      author={Yaolun Zhang and Xiaogeng Liu and Chaowei Xiao},
      year={2025},
      eprint={2507.22606},
      archivePrefix={arXiv},
      primaryClass={cs.AI},
      url={https://arxiv.org/abs/2507.22606}, 
}

@misc{han2025mapgdmultiagentpromptgradient,
      title={MAPGD: Multi-Agent Prompt Gradient Descent for Collaborative Prompt Optimization}, 
      author={Yichen Han and Bojun Liu and Zhengpeng zhou and Guanyu Liu and Zeng Zhang and Yang Yang and Wenli Wang and Isaac N Shi and Yunyan and Lewei He and Tianyu Shi},
      year={2025},
      eprint={2509.11361},
      archivePrefix={arXiv},
      primaryClass={cs.AI},
      url={https://arxiv.org/abs/2509.11361}, 
}

@misc{zhang2025agentic-supernet,
      title={Multi-agent Architecture Search via Agentic Supernet}, 
      author={Guibin Zhang and Luyang Niu and Junfeng Fang and Kun Wang and Lei Bai and Xiang Wang},
      year={2025},
      eprint={2502.04180},
      archivePrefix={arXiv},
      primaryClass={cs.LG},
      url={https://arxiv.org/abs/2502.04180}, 
}

@misc{tang2025hivaselforganizedhierarchicalvariable,
      title={HiVA: Self-organized Hierarchical Variable Agent via Goal-driven Semantic-Topological Evolution}, 
      author={Jinzhou Tang and Jusheng Zhang and Qinhan Lv and Sidi Liu and Jing Yang and Chengpei Tang and Keze Wang},
      year={2025},
      eprint={2509.00189},
      archivePrefix={arXiv},
      primaryClass={cs.AI},
      url={https://arxiv.org/abs/2509.00189}, 
}

@inproceedings{chen2025smurfs,
  title={Smurfs: Multi-Agent System using Context-Efficient DFSDT for Tool Planning},
  author={Chen, Junzhi and Liang, Juhao and Wang, Benyou},
  booktitle={Proceedings of the 2025 Conference of the Nations of the Americas Chapter of the Association for Computational Linguistics: Human Language Technologies (Volume 1: Long Papers)},
  pages={3281--3298},
  year={2025}
}

@misc{mou2025ecolangefficienteffectiveagent,
      title={EcoLANG: Efficient and Effective Agent Communication Language Induction for Social Simulation}, 
      author={Xinyi Mou and Chen Qian and Wei Liu and Xuanjing Huang and Zhongyu Wei},
      year={2025},
      eprint={2505.06904},
      archivePrefix={arXiv},
      primaryClass={cs.CL},
      url={https://arxiv.org/abs/2505.06904}, 
}

@misc{yao2023reactsynergizingreasoningacting,
      title={ReAct: Synergizing Reasoning and Acting in Language Models}, 
      author={Shunyu Yao and Jeffrey Zhao and Dian Yu and Nan Du and Izhak Shafran and Karthik Narasimhan and Yuan Cao},
      year={2023},
      eprint={2210.03629},
      archivePrefix={arXiv},
      primaryClass={cs.CL},
      url={https://arxiv.org/abs/2210.03629}, 
}

@misc{qian2025smartselfawareagenttool,
      title={SMART: Self-Aware Agent for Tool Overuse Mitigation}, 
      author={Cheng Qian and Emre Can Acikgoz and Hongru Wang and Xiusi Chen and Avirup Sil and Dilek Hakkani-Tür and Gokhan Tur and Heng Ji},
      year={2025},
      eprint={2502.11435},
      archivePrefix={arXiv},
      primaryClass={cs.AI},
      url={https://arxiv.org/abs/2502.11435}, 
}

@misc{xiong2025memorymanagementimpactsllm,
      title={How Memory Management Impacts LLM Agents: An Empirical Study of Experience-Following Behavior}, 
      author={Zidi Xiong and Yuping Lin and Wenya Xie and Pengfei He and Jiliang Tang and Himabindu Lakkaraju and Zhen Xiang},
      year={2025},
      eprint={2505.16067},
      archivePrefix={arXiv},
      primaryClass={cs.AI},
      url={https://arxiv.org/abs/2505.16067}, 
}

@article{gao2022pal,
  title={PAL: Program-aided Language Models},
  author={Gao, Luyu and Madaan, Aman and Zhou, Shuyan and Alon, Uri and Liu, Pengfei and Yang, Yiming and Callan, Jamie and Neubig, Graham},
  journal={arXiv preprint arXiv:2211.10435},
  year={2022}
}

@misc{chen2021evaluatinglargelanguagemodels,
      title={Evaluating Large Language Models Trained on Code}, 
      author={Mark Chen and Jerry Tworek and Heewoo Jun and Qiming Yuan and Henrique Ponde de Oliveira Pinto and Jared Kaplan and Harri Edwards and Yuri Burda and Nicholas Joseph and Greg Brockman and Alex Ray and Raul Puri and Gretchen Krueger and Michael Petrov and Heidy Khlaaf and Girish Sastry and Pamela Mishkin and Brooke Chan and Scott Gray and Nick Ryder and Mikhail Pavlov and Alethea Power and Lukasz Kaiser and Mohammad Bavarian and Clemens Winter and Philippe Tillet and Felipe Petroski Such and Dave Cummings and Matthias Plappert and Fotios Chantzis and Elizabeth Barnes and Ariel Herbert-Voss and William Hebgen Guss and Alex Nichol and Alex Paino and Nikolas Tezak and Jie Tang and Igor Babuschkin and Suchir Balaji and Shantanu Jain and William Saunders and Christopher Hesse and Andrew N. Carr and Jan Leike and Josh Achiam and Vedant Misra and Evan Morikawa and Alec Radford and Matthew Knight and Miles Brundage and Mira Murati and Katie Mayer and Peter Welinder and Bob McGrew and Dario Amodei and Sam McCandlish and Ilya Sutskever and Wojciech Zaremba},
      year={2021},
      eprint={2107.03374},
      archivePrefix={arXiv},
      primaryClass={cs.LG},
      url={https://arxiv.org/abs/2107.03374}, 
}

@misc{austin2021programsynthesislargelanguage,
      title={Program Synthesis with Large Language Models}, 
      author={Jacob Austin and Augustus Odena and Maxwell Nye and Maarten Bosma and Henryk Michalewski and David Dohan and Ellen Jiang and Carrie Cai and Michael Terry and Quoc Le and Charles Sutton},
      year={2021},
      eprint={2108.07732},
      archivePrefix={arXiv},
      primaryClass={cs.PL},
      url={https://arxiv.org/abs/2108.07732}, 
}

@inproceedings{dua-etal-2019-drop,
    title = "{DROP}: A Reading Comprehension Benchmark Requiring Discrete Reasoning Over Paragraphs",
    author = "Dua, Dheeru  and
      Wang, Yizhong  and
      Dasigi, Pradeep  and
      Stanovsky, Gabriel  and
      Singh, Sameer  and
      Gardner, Matt",
    editor = "Burstein, Jill  and
      Doran, Christy  and
      Solorio, Thamar",
    booktitle = "Proceedings of the 2019 Conference of the North {A}merican Chapter of the Association for Computational Linguistics: Human Language Technologies, Volume 1 (Long and Short Papers)",
    month = jun,
    year = "2019",
    address = "Minneapolis, Minnesota",
    publisher = "Association for Computational Linguistics",
    url = "https://aclanthology.org/N19-1246/",
    doi = "10.18653/v1/N19-1246",
    pages = "2368--2378"
}

@misc{HuggingFaceH4_AIME_2024,
  author       = {{HuggingFaceH4}},
  title        = {AIME 2024 Dataset},
  year         = {2024},
  publisher    = {Hugging Face},
  howpublished = {\url{https://huggingface.co/datasets/HuggingFaceH4/aime_2024}}
}

@misc{yu2025aworldorchestratingtrainingrecipe,
      title={AWorld: Orchestrating the Training Recipe for Agentic AI}, 
      author={Chengyue Yu and Siyuan Lu and Chenyi Zhuang and Dong Wang and Qintong Wu and Zongyue Li and Runsheng Gan and Chunfeng Wang and Siqi Hou and Gaochi Huang and Wenlong Yan and Lifeng Hong and Aohui Xue and Yanfeng Wang and Jinjie Gu and David Tsai and Tao Lin},
      year={2025},
      eprint={2508.20404},
      archivePrefix={arXiv},
      primaryClass={cs.AI},
      url={https://arxiv.org/abs/2508.20404}, 
}

@misc{wang2023selfconsistencyimproveschainthought,
      title={Self-Consistency Improves Chain of Thought Reasoning in Language Models}, 
      author={Xuezhi Wang and Jason Wei and Dale Schuurmans and Quoc Le and Ed Chi and Sharan Narang and Aakanksha Chowdhery and Denny Zhou},
      year={2023},
      eprint={2203.11171},
      archivePrefix={arXiv},
      primaryClass={cs.CL},
      url={https://arxiv.org/abs/2203.11171}, 
}

@misc{hu2025owl,
      title={OWL: Optimized Workforce Learning for General Multi-Agent Assistance in Real-World Task Automation}, 
      author={Mengkang Hu and Yuhang Zhou and Wendong Fan and Yuzhou Nie and Bowei Xia and Tao Sun and Ziyu Ye and Zhaoxuan Jin and Yingru Li and Qiguang Chen and Zeyu Zhang and Yifeng Wang and Qianshuo Ye and Bernard Ghanem and Ping Luo and Guohao Li},
      year={2025},
      eprint={2505.23885},
      archivePrefix={arXiv},
      primaryClass={cs.AI},
      url={https://arxiv.org/abs/2505.23885}, 
}

@misc{openai_gpt4.1_2025,
  author       = {OpenAI},
  title        = {Introducing GPT-4.1 in the API},
  howpublished = {OpenAI blog post},
  year         = {2025},
  month        = {Apr},
  day          = {14},
  url          = {https://openai.com/index/gpt-4-1/}
}

@misc{comanici2025gemini25pushingfrontier,
      title={Gemini 2.5: Pushing the Frontier with Advanced Reasoning, Multimodality, Long Context, and Next Generation Agentic Capabilities}, 
      author={Gemini Team},
      year={2025},
      eprint={2507.06261},
      archivePrefix={arXiv},
      primaryClass={cs.CL},
      url={https://arxiv.org/abs/2507.06261}, 
}

@misc{qwen3technicalreport,
      title={Qwen3 Technical Report}, 
      author={Qwen Team},
      year={2025},
      eprint={2505.09388},
      archivePrefix={arXiv},
      primaryClass={cs.CL},
      url={https://arxiv.org/abs/2505.09388}, 
}

@misc{zhang2025aflowautomatingagenticworkflow,
      title={AFlow: Automating Agentic Workflow Generation}, 
      author={Jiayi Zhang and Jinyu Xiang and Zhaoyang Yu and Fengwei Teng and Xionghui Chen and Jiaqi Chen and Mingchen Zhuge and Xin Cheng and Sirui Hong and Jinlin Wang and Bingnan Zheng and Bang Liu and Yuyu Luo and Chenglin Wu},
      year={2025},
      eprint={2410.10762},
      archivePrefix={arXiv},
      primaryClass={cs.AI},
      url={https://arxiv.org/abs/2410.10762}, 
}

@article{cobbe2021gsm8k,
  title={Training Verifiers to Solve Math Word Problems},
  author={Cobbe, Karl and Kosaraju, Vineet and Bavarian, Mohammad and Chen, Mark and Jun, Heewoo and Kaiser, Lukasz and Plappert, Matthias and Tworek, Jerry and Hilton, Jacob and Nakano, Reiichiro and Hesse, Christopher and Schulman, John},
  journal={arXiv preprint arXiv:2110.14168},
  year={2021}
}

@misc{xie2025profileawaremaneuveringdynamicmultiagent,
      title={Profile-Aware Maneuvering: A Dynamic Multi-Agent System for Robust GAIA Problem Solving by AWorld}, 
      author={Zhitian Xie and Qintong Wu and Chengyue Yu and Chenyi Zhuang and Jinjie Gu},
      year={2025},
      eprint={2508.09889},
      archivePrefix={arXiv},
      primaryClass={cs.AI},
      url={https://arxiv.org/abs/2508.09889}, 
}
\bibliographystyle{configuration/iclr2026_conference}

\clearpage
\appendix
\onecolumn
{
    \hypersetup{linkcolor=black}
    \parskip=0em
    \renewcommand{\contentsname}{Contents}
    \tableofcontents
    \addtocontents{toc}{\protect\setcounter{tocdepth}{3}}
}

\newpage

\section{Appendix}
\subsection{LLM usage}
The large language model (LLM) was utilized as a writing assistant during the preparation of this manuscript. Its application was strictly limited to improving the clarity and grammatical accuracy of the text. Specific uses included rephrasing sentences for better flow and translating initial concepts and drafts from Chinese to English. All core scientific contributions, including the conceptualization of our \method framework, the design of the methodology and experiments, and the analysis and interpretation of the results, are solely the work of the authors. The authors take full responsibility for all claims and the final content of this paper.

\subsection{Experimental Setup}
\label{experiment_details}
\subsubsection{Datasets}

Here, we provide a detailed introduction to the datasets used in this paper:

\begin{itemize}[left=0pt]
    \item \textbf{GAIA}~\citep{mialon2023gaiabenchmarkgeneralai} serves as a benchmark designed to evaluate next-generation LLMs that possess enhanced capabilities through the incorporation of tools, efficient prompting strategies, and access to external search resources. This benchmark comprises over 450 challenging questions, each with a clear and unequivocal answer, necessitating varying degrees of tooling and autonomy for resolution. Accordingly, the questions are categorized into three distinct levels: Level 1 is expected to be solvable by proficient LLMs, while Level 3 signifies a substantial increase in the model's capabilities. Each level includes a fully public development set for validation purposes, as well as a test set containing private answers and associated metadata. In our experiments, we utilize the test set, which encompasses 164 tasks.
    \item \textbf{GSM-hard}~\citep{gao2022pal} is an advanced version of the GSM8K mathematics reasoning dataset~\citep{cobbe2021gsm8k}. This enhanced dataset presents models with increased challenges, featuring larger numerical values and more complex relationships within the problems.
    \item \textbf{AIME-2024}~\citep{HuggingFaceH4_AIME_2024} is a dataset comprising problems derived from the American Invitational Mathematics Examination (AIME) 2024. AIME is a prestigious mathematics competition for high school students, recognized for its challenging problems that span various mathematical domains. This benchmark serves multiple purposes: it evaluates the mathematical reasoning capabilities of LLMs, assesses their problem-solving abilities on complex mathematical challenges, and investigates AI performance on structured mathematical tasks.
    \item \textbf{HumanEval}~\citep{chen2021evaluatinglargelanguagemodels} is a dataset released by OpenAI that includes 164 programming problems, each containing a function signature, a docstring, a body, and several associated unit tests. These problems were handwritten to ensure that they were not included in the training dataset for code-generation models. This benchmark is crucial for evaluating code-generation models, providing a structured set of challenges in Python that facilitates the assessment of both the quality and correctness of code produced by language models.
    \item \textbf{MBPP}(Mostly Basic Python Problems Dataset)~\citep{austin2021programsynthesislargelanguage} comprises approximately 1,000 crowd-sourced Python programming problems that are specifically designed to be solvable by entry-level programmers. The dataset covers essential programming fundamentals and standard library functionalities. Each problem includes a task description, a corresponding code solution, and three automated test cases.
    \item \textbf{DROP}(Data Retrieval Open Answering)~\cite{dua-etal-2019-drop} is a reading comprehension benchmark that requires discrete reasoning over paragraphs. This dataset consists of 96,000 questions developed through crowd sourcing and adversarial methods. It challenges systems to resolve references within the questions, which may point to multiple input positions. The tasks entail performing discrete operations, such as addition, counting, and sorting, necessitating a substantially more comprehensive understanding of paragraph content than that demanded by prior datasets. In our experiment, we sampled 800 tasks for evaluation.
\end{itemize}

\subsubsection{Baselines}
\label{appendix:baselines}
\begin{itemize}[left=0pt]
    \item \textbf{Vanilla} is the original Large Language Model (LLM) that processes input using only the question and a basic prompt, without any prompt engineering or external tool integration. This straightforward approach emphasizes the model's inherent capabilities in handling natural language tasks. By operating in this simplistic manner, Vanilla LLM serves as a critical baseline for evaluating the performance of more advanced techniques that incorporate sophisticated prompt strategies or additional tools, thereby providing valuable insights into the effectiveness of various methodologies in natural language processing.
    \item \textbf{CoT-SC}(Chain-of-Thought Self-Consistency)~\citep{wang2023selfconsistencyimproveschainthought} serves as a baseline for enhancing the reasoning capabilities of language models. This approach generates multiple reasoning chains, which are then aggregated to produce a coherent summary. By leveraging self-consistency, CoT-SC improves the reliability of the model's outputs, allowing for better performance in complex reasoning tasks. This structured process facilitates deeper analysis of the model's thought processes, providing a foundation for comparing more advanced reasoning strategies and understanding their impact on overall performance.
    \item \textbf{MetaAgent}~\citep{zhang2025metaagentautomaticallyconstructingmultiagent} is a groundbreaking framework designed to automatically construct multi-agent systems by specifying the objectives of a given task. A distinctive feature of MetaAgent is its ability to generate these multi-agent systems without relying on external training data. This capability allows the produced multi-agent systems to effectively address all scenarios within the specified task domain. The underlying architecture of the Multi-Agent System is based on Finite State Machines(FSM), which facilitates structured decision-making and state transitions, thereby enhancing the system's operational efficiency and adaptability.
    \item \textbf{OWL}(Open Web Language)~\citep{hu2025owl} serves as a foundational framework for knowledge representation in multi-agent systems. By enabling agents to process and reason over complex data in a machine-readable format, OWL is crucial for facilitating interoperability among diverse agents. It allows for the creation of ontologies that define intricate relationships and constraints within the environment, thereby enhancing collaborative behaviors among agents. The expressive power of OWL supports advanced inference capabilities, empowering agents to share knowledge effectively and make informed decisions. This framework establishes a robust baseline for evaluating and enhancing the performance of multi-agent systems in various applications.
    \item \textbf{Smolagent}~\citep{smolagents} is a lightweight library designed to facilitate the development and implementation of AI agents that can think and operate using code. It emphasizes simplicity and efficiency, enabling users to create multi-agent systems with minimal code. Smolagent's architecture allows for smart threading, dependency management, and context sharing, making it ideal for orchestrating complex tasks. By providing a streamlined framework, Smolagent serves as a foundational model for evaluating the performance and capabilities of more advanced agent-based systems in various applications.
    \item \textbf{OAgents}~\citep{zhu2025oagentsempiricalstudybuilding} is a modular multi-agent framework that conducts a thorough empirical study of key agent components (planning, memory, tool use, test-time scaling) on benchmarks such as GAIA and BrowseComp. It delivers great performance among open-source agent frameworks. Importantly, OAgents builds on the lightweight agentience model provided by Smolagent (which emphasises code-based agent orchestration and minimal overhead) and extends it with fine-grained task decomposition, dynamic workflow adaptation, multi-source web browsing and more extensive tool and memory modules.
    \item \textbf{AWorld}~\citep{yu2025aworldorchestratingtrainingrecipe} is an open-source framework for large-scale agent–environment interaction, designed to operationalize the “learning from practice” paradigm in agentic AI. It features a hierarchical multi-agent architecture composed of specialized agents such as the \emph{Execution Agent}, which performs primary reasoning and tool-use operations, and the \emph{Guard Agent}, which intervenes at critical steps to verify and refine intermediate outcomes. AWorld adopts a modular design supporting dynamic supervision, context tracking, and distributed orchestration, enabling efficient coordination across diverse tasks and environments. By treating agents and tools as interchangeable components within a unified orchestration layer, it facilitates flexible composition, concurrent execution, and fine-grained control over reasoning workflows, illustrating a scalable and extensible paradigm for constructing adaptive multi-agent systems.
\end{itemize}

\subsection{Implementation details}
\label{implementation_details}

\begin{table*}[t]
    \centering
    \caption{Hyperparameter settings for the Heuristic-Based Adaptive Filter across different benchmarks. The symbols correspond to the definitions in Algorithm \ref{alg:adaptive_filter}.}
    \label{tab:hyperparams}
    \vspace{2mm} %
    \resizebox{\textwidth}{!}{%
    \begin{tabular}{llcccccc}
    \toprule
    \textbf{Condition} & \textbf{Parameter (Symbol)} & \textbf{GAIA} & \textbf{HumanEval} & \textbf{MBPP} & \textbf{AIME} & \textbf{DROP} & \textbf{GSM-Hard} \\
    \midrule
    \multirow{2}{*}{\textit{Inefficient}} 
        & Step Check Interval ($\tau_{\text{step}}$) & 8 & 6 & 4 & 4 & 4 & 4 \\
        & Loop Detection Window ($\tau_{\text{loop}}$) & 5 & 5 & 3 & 3 & 3 & 3 \\
    \midrule
    \textit{Excessive} 
        & Length Threshold ($\tau_{\text{len}}$) & \multicolumn{6}{c}{3000} \\
    \bottomrule
    \end{tabular}
    }
\end{table*}

In this section, we provide a detailed description of how the conceptual framework of \method{} is implemented in our codebase. Our implementation is centered around the \texttt{supervise\_and\_correct} function, which serves as the primary entry point for all supervisory actions. We structure our explanation following the same \emph{What, When, and How} logic presented in our main methodology.

\subsubsection{What to Supervise: The ActionStep Object}
Our supervision targets the discrete interaction steps performed by each agent within the MAS. In our framework, every such interaction is encapsulated in a data structure we refer to as an \texttt{ActionStep} object. This object contains all relevant information for a single step, including the agent's thought process (\texttt{model\_output}), the executed \texttt{tool\_calls}, the resulting \texttt{observations}, and an \texttt{error} attribute which is populated if an exception occurs. Our \method{} is implemented as a callback function that intercepts every \texttt{ActionStep} object generated by any agent in the system.

\subsubsection{When to Supervise: The Prioritized Adaptive Filter}
To avoid the prohibitive cost of constant intervention, we employ a lightweight, LLM-free adaptive filter. This filter is implemented as a prioritized conditional chain at the beginning of the \texttt{supervise\_and\_correct} function. It evaluates each \texttt{ActionStep} to determine if supervision is warranted. The conditions are checked in the following order of precedence:

\begin{enumerate}[nosep, leftmargin=15pt]
    \item \textbf{Sub-Agent Completion}: The highest priority is to check if the observation contains a final report from a sub-agent (identified by the presence of a \texttt{"\textless summary\_of\_work\textgreater"} string). If so, it triggers the specialized \texttt{Adaptive Observation Purification} strategy to distill the findings for the manager agent.
    \item \textbf{Error Occurrence}: If the \texttt{step.error} attribute is not \texttt{None}, the \texttt{Proactive Error Correction} strategy is triggered. Our implementation includes a defensive check to ensure this does not fire for known, non-critical tool failures that the base agent can handle.
    \item \textbf{Inefficient Behavior}: If no error is present, we then check for inefficiency using our heuristic-based \texttt{\_check\_for\_inefficiency} function. This function detects patterns such as hard loops (identical actions and observations) and excessive step counts for a given sub-task, triggering the \texttt{Guidance for Inefficiency} strategy.
    \item \textbf{Excessive Observation Length}: Finally, if none of the above conditions are met, the filter checks if the length of the \texttt{step.observations} string exceeds a pre-defined threshold $\tau_{\text{len}}$ (3,000 characters in our implementation). If it does, the general type of \texttt{Adaptive Observation Purification} strategy is activated.
\end{enumerate}

If none of these trigger conditions are met, the step is approved by default, thereby avoiding any unnecessary LLM-based supervision overhead. 
The filter’s sensitivity is governed by three key hyperparameters: $\tau_{\text{step}}$ and $\tau_{\text{loop}}$ modulate the detection of inefficient behaviors, while $\tau_{\text{len}}$ defines the threshold for identifying excessive observations.
The complete logic of this heuristic-based mechanism is formalized in Algorithm \ref{alg:adaptive_filter}.

\begin{algorithm}[t]
\caption{Heuristic-Based Adaptive Filter Logic}
\label{alg:adaptive_filter}
\begin{algorithmic}[1]
\Input Current execution history $\mathcal{H}$ (sequence of steps), Current observation $o$, Error status $e$
\Require Hyperparameters: $\tau_{\text{step}}$ (step check interval), $\tau_{\text{loop}}$ (loop detection window), $\tau_{\text{len}}$ (length threshold)
\Ensure Boolean flag $trigger$, Intervention context $c$

\State $trigger \gets \text{False}$
\State $c \gets \text{None}$

\Statex \Comment{\textit{Priority 1: Check for explicit runtime errors}}
\If{$e$ is \textbf{True}}
    \State \Return $(\text{True}, c_{\text{error}})$
\EndIf

\Statex \Comment{\textit{Priority 2: Check for inefficient behaviors}}
\State $N \gets \text{Length}(\mathcal{H})$
\If{$N > 0$ \textbf{and} $N \pmod{\tau_{\text{step}}} = 0$} \Comment{Periodic strategy check}
    \State \Return $(\text{True}, c_{\text{inefficient}})$
\EndIf

\If{$N \geq \tau_{\text{loop}}$} \Comment{Repetitive loop detection}
    \State $\mathcal{A}_{\text{recent}} \gets \text{GetLastToolCalls}(\mathcal{H}, \text{window}=\tau_{\text{loop}})$
    \If{$|\text{Unique}(\mathcal{A}_{\text{recent}})| = 1$}
        \State \Return $(\text{True}, c_{\text{inefficient}})$
    \EndIf
\EndIf

\Statex \Comment{\textit{Priority 3: Check for excessive information}}
\If{$\text{Length}(o) > \tau_{\text{len}}$}
    \State \Return $(\text{True}, c_{\text{excessive}})$
\EndIf

\State \Return $($trigger, $c)$
\end{algorithmic}
\end{algorithm}

\begin{figure}[!t] %
    \centering
    \includegraphics[width=0.9\linewidth]{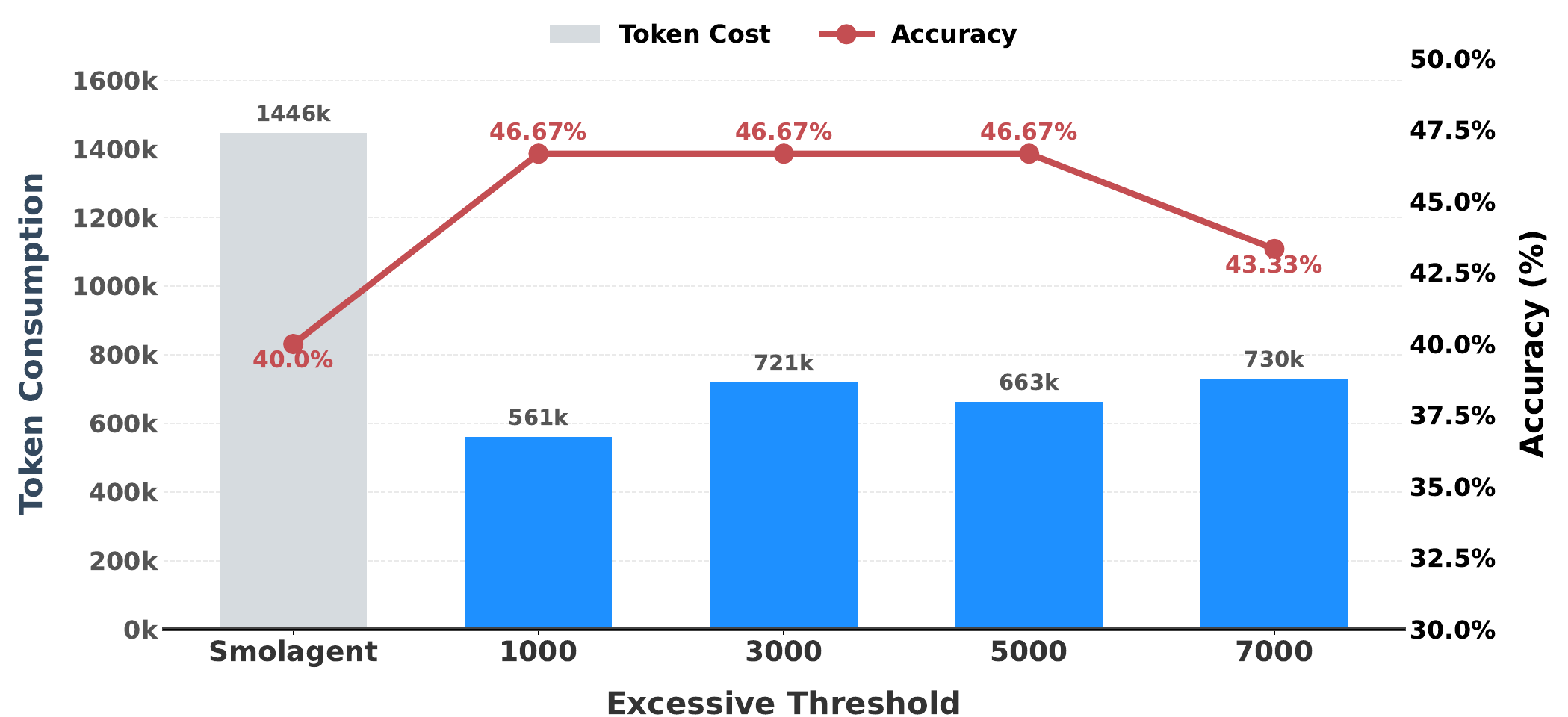}
    \caption{
        \textbf{Sensitivity analysis on excessive threshold ($\tau_{\text{len}}$).} Evaluated on GAIA subset (top-10 most token-intensive tasks per level).
    }
    \label{fig:sensitivity}
\end{figure}

\paragraph{Sensitivity analysis and hyperparameter configuration.}
Table \ref{tab:hyperparams} details the hyperparameter settings for our Heuristic-Based Adaptive Filter across different benchmarks. We conducted a sensitivity analysis on the excessive observation length threshold ($\tau_{\text{len}}$) using the representative subset of the GAIA benchmark (top-10 most token-intensive tasks per level), as illustrated in Figure \ref{fig:sensitivity}. The results indicate that our method maintains robust performance across a wide range of threshold values, demonstrating its adaptability without significant degradation. Although slight performance peaks are observed at $\tau_{\text{len}}=1000$ and $\tau_{\text{len}}=5000$, we selected $3000$ as the default value. This choice strikes a prudent balance between sensitivity (catching enough noise) and specificity (preserving useful context), ensuring optimal performance across diverse tasks while minimizing the risk of over-intervention.

\paragraph{Configuration for OAgents.}
For the OAgents framework~\citep{zhu2025oagentsempiricalstudybuilding}, we adjusted the parameters to $\tau_{\text{step}}=6$, $\tau_{\text{loop}}=3$, and $\tau_{\text{len}}=10000$. The significantly higher $\tau_{\text{len}}$ (compared to 3000 in Smolagent) is necessitated by OAgents' architecture, which tends to generate extensive verbose outputs due to its complex tool usage and memory retrieval modules. A higher threshold is essential here to effectively identify truly excessive information without triggering false positives on standard OAgents operations.

\paragraph{Adaptation for AWorld: An MCP-Based Approach.}
Unlike the external heuristic filter used in Smolagent, the integration with AWorld~\citep{yu2025aworldorchestratingtrainingrecipe} leverages its native tool-use architecture. We implemented the \method as a \textbf{Model Context Protocol (MCP)} service, allowing it to be dynamically discovered and invoked by the AWorld agent. This adaptation involves three key modifications:
\begin{itemize}[nosep, leftmargin=12pt]
    \item \textbf{Mandatory Invocation via System Prompt:} We refined AWorld's system prompt to enforce a protocol where the agent \textit{must} invoke the \method during critical phases—specifically ``Information Gathering'' and ``Thinking Process Reviewing''. This ensures the \method acts as a mandatory gatekeeper for logical consistency.
    \item \textbf{Capability-Based Routing:} We defined a specific MCP schema where the \method broadcasts its capabilities, including \texttt{Error Root Cause Diagnosis}, \texttt{Structured Information Synthesis}, and \texttt{Workflow Efficiency Assessment}. This allows the AWorld agent to match its current execution status (e.g., encountering an exception or synthesizing search results) with the appropriate Supervisor function.
    \item \textbf{Trigger Mechanism:} Instead of counting token length, the trigger is semantic. Explicit error returns from AWorld's system serve as triggers for \texttt{error\_analysis}, while the \texttt{sub\_agent\_result\_synthesis} and \texttt{inefficiency\_analysis} modes are seamlessly integrated into the MCP process flow to facilitate output verification.
\end{itemize}

\subsubsection{How to Supervise: The Intervention Pipeline}
Once the adaptive filter flags an interaction, the \texttt{supervise\_and\_correct} function executes a three-stage intervention pipeline:

\paragraph{1. Context Aggregation}
Before making a decision, the Supervisor aggregates a context window ($\mathcal{W}$). This process involves retrieving the global task ($G$) and the agent's local task ($L$), formatting the agent's recent local action history ($T_l$) via the \texttt{\_format\_local\_trace\_for\_prompt} function, and generating a summary of the current step ($S$) using the \texttt{\_summarize\_interaction} function. For inefficient behavior, the full global trace ($T_g$) is also included.

\paragraph{2. LLM-based Decision Making}
The aggregated context is then compiled into a specialized prompt tailored to the triggered supervision type (e.g., \texttt{Proactive Error Correction}). This prompt instructs our main model (e.g., GPT-4.1) to analyze the situation and return a structured JSON object containing its \texttt{analysis}, a chosen \texttt{action} (from the set \{\texttt{approve}, \texttt{correct\_observation}, \texttt{provide\_guidance}, \texttt{run\_verification}\}), and the necessary \texttt{parameters} to execute that action.

\paragraph{3. Action Execution}
The returned JSON is parsed, and the chosen action is executed.
\begin{itemize}[nosep, leftmargin=12pt]
    \item \texttt{correct\_observation}: The original \texttt{step.observations} is entirely replaced with the \texttt{new\_observation} provided in the parameters. A ``[Supervisor's Note: ...]'' is prepended to inform the agent of the modification.
    \item \texttt{provide\_guidance}: The \texttt{guidance} string from the parameters is appended to the end of the existing \texttt{step.observations}, leaving the original sensory data intact while providing a corrective hint.
    \item \texttt{run\_verification}: The \texttt{task} parameter is passed to a dedicated, fully-equipped verification agent, and its conclusive findings are appended to the \texttt{step.observations}.
\end{itemize}

\subsection{Extended Experimental Analysis}

\subsubsection{Overhead analysis of \method}
\label{overhead_analysis}

\begin{figure}[!t] %
    \centering
    \includegraphics[width=0.9\linewidth]{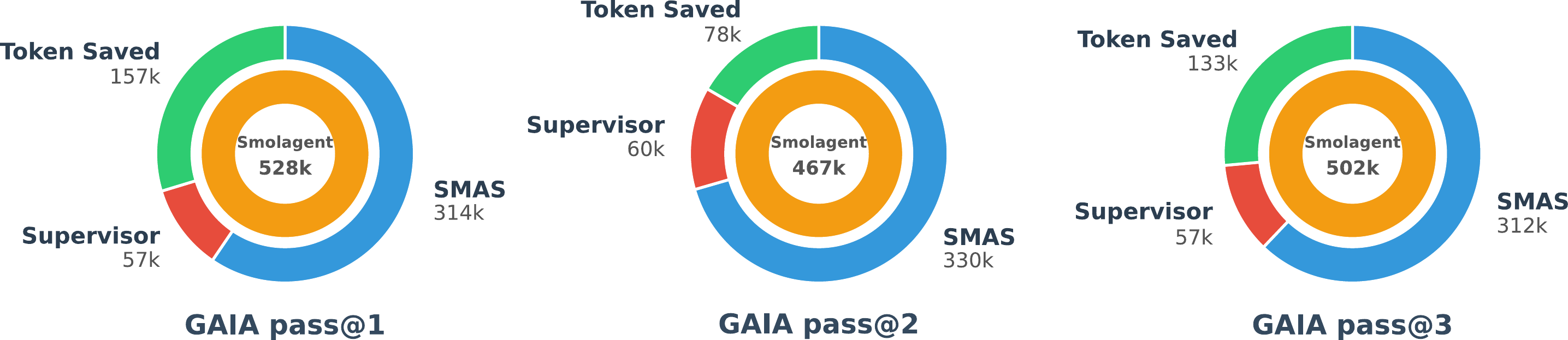}
    \vspace{-0.5em}
    \caption{
        \textbf{\method overhead on the GAIA benchmark.}
    }
    \label{fig:super_overhead}
\end{figure}

\paragraph{Token overhead analysis}
The token overhead of \method itself is shown in Table~\ref{tab:super_overhead} and Figure~\ref{fig:super_overhead}. We analyze the token consumption on the GAIA validation set under different pass@k settings. The results indicate that integrating \method leads to a significant reduction in overall token usage across all complexity levels (L1, L2, L3) and pass@k configurations. Specifically, \method achieves an average token saving of \textbf{35.95\%} across all settings compared to the baseline Smolagent. This substantial decrease in token consumption highlights \method's effectiveness in optimizing the multi-agent system's efficiency by reducing unnecessary interactions and streamlining the reasoning process.

Notably, even when accounting for the additional tokens introduced by \method's supervisory interventions, the net token consumption remains significantly lower than that of the baseline (already reported in main content). 
And the token overhead of \method only contains about \textbf{15.45\%} in average of the total tokens used in the Smolagent baseline.
This demonstrates that the benefits of improved efficiency and reduced redundancy far outweigh the costs associated with supervision.

\begin{table}[!t]
    \caption{
        \textbf{Token efficiency analysis on GAIA validation set.}
        Comparison of token consumption across different pass@k settings.
    }
    \label{tab:super_overhead}
    \centering
    \resizebox{\textwidth}{!}{
        \begin{tabular}{@{}lcccc@{}}
            \toprule
            \multicolumn{1}{c}{\bf Method} & \multicolumn{1}{c}{\bf Avg. Tokens (K)} & \multicolumn{1}{c}{\bf L1 Tokens (K)} & \multicolumn{1}{c}{\bf L2 Tokens (K)} & \multicolumn{1}{c}{\bf L3 Tokens (K)} \\
            \midrule
            \multicolumn{5}{c}{\bf pass@1} \\
            \midrule
            Smolagent & 527.76 & 298.51 & 619.59 & 691.33 \\
            \, + Supervised MAS & 314.07 {\scriptsize \textcolor{blue}{↓40.49\%}} & 220.63 {\scriptsize \textcolor{blue}{↓26.09\%}} & 342.18 {\scriptsize \textcolor{blue}{↓44.77\%}} & 411.58 {\scriptsize \textcolor{blue}{↓40.47\%}} \\
            \, + Supervised MAS (NET) & 371.12 {\scriptsize \textcolor{blue}{↓29.68\%}} & 258.28 {\scriptsize \textcolor{blue}{↓13.48\%}} & 404.96 {\scriptsize \textcolor{blue}{↓34.64\%}} & 489.22 {\scriptsize \textcolor{blue}{↓29.23\%}} \\
            \midrule
            \multicolumn{5}{c}{\bf pass@2} \\
            \midrule
            Smolagent & 467.19 & 275.85 & 548.02 & 589.92 \\
            \, + Supervised MAS & 329.51 {\scriptsize \textcolor{blue}{↓29.47\%}} & 231.96 {\scriptsize \textcolor{blue}{↓15.91\%}} & 354.21 {\scriptsize \textcolor{blue}{↓35.37\%}} & 446.64 {\scriptsize \textcolor{blue}{↓24.29\%}} \\
            \, + Supervised MAS (NET) & 389.55 {\scriptsize \textcolor{blue}{↓16.62\%}} & 270.07 {\scriptsize \textcolor{blue}{↓2.10\%}} & 420.97 {\scriptsize \textcolor{blue}{↓23.18\%}} & 529.20 {\scriptsize \textcolor{blue}{↓10.29\%}} \\
            \midrule
            \multicolumn{5}{c}{\bf pass@3} \\
            \midrule
            Smolagent & 502.40 & 282.14 & 605.05 & 611.87 \\
            \, + Supervised MAS & 312.06 {\scriptsize \textcolor{blue}{↓37.89\%}} & 236.28 {\scriptsize \textcolor{blue}{↓16.25\%}} & 342.36 {\scriptsize \textcolor{blue}{↓43.42\%}} & 366.31 {\scriptsize \textcolor{blue}{↓40.13\%}} \\
            \, + Supervised MAS (NET) & 369.52 {\scriptsize \textcolor{blue}{↓26.45\%}} & 276.84 {\scriptsize \textcolor{blue}{↓1.88\%}} & 409.05 {\scriptsize \textcolor{blue}{↓32.39\%}} & 427.72 {\scriptsize \textcolor{blue}{↓30.10\%}} \\
            \bottomrule
        \end{tabular}
    }
\end{table}

\begin{figure}[!t] %
    \centering
    \includegraphics[width=0.9\linewidth]{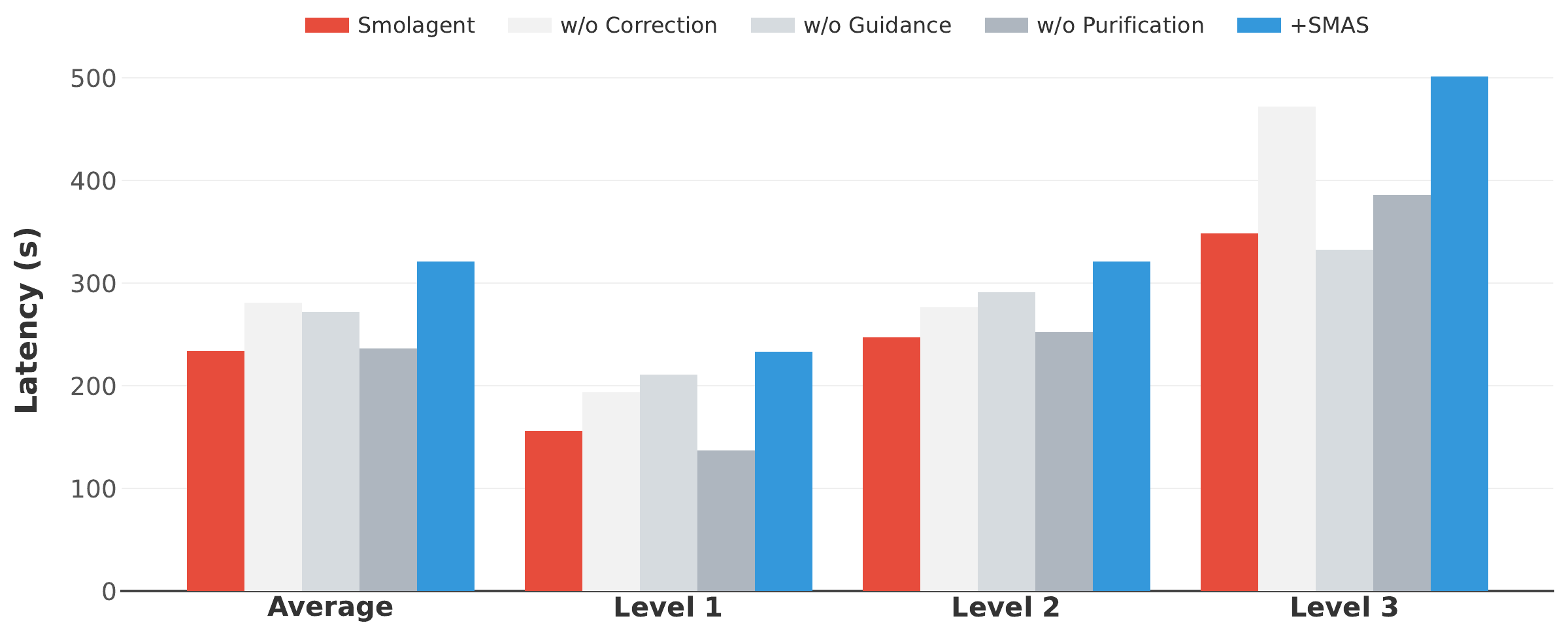}
    \vspace{-0.5em}
    \caption{
        \textbf{\method latency on the GAIA benchmark.}
    }
    \label{fig:super_latency}
\end{figure}

\paragraph{Latency Overhead Analysis.}
Table \ref{tab:ablation_latency} and Figure \ref{fig:super_latency} analyzes the temporal impact of \method. While integrating the supervisor introduces an average latency increase of \textbf{37.27\%}, this translates to an absolute delay of less than \textbf{1.5 minutes} per task. Crucially, the ablation study reveals that Adaptive Observation Purification is the primary driver of this latency. Notably, the \textit{w/o Purification} variant exhibits a runtime nearly identical to the baseline (236.21s vs. 233.96s). This indicates that the overhead is strictly tied to the processing of excessive information. We consider this a strategic trade-off: exchanging a modest temporal cost for substantial economic (token) savings and enhanced system robustness.

\begin{table}[!t]
    \caption{
        \textbf{Ablation study of \method's components regarding Latency} on the GAIA validation set.
        Average latency (in seconds) is reported for different complexity levels.
    }
    \label{tab:ablation_latency}
    \centering
    \resizebox{\textwidth}{!}{
        \begin{tabular}{>{\raggedright\arraybackslash}m{4cm} c c c c c}
            \toprule
            \bf Method                   & \bf Avg. Acc. & \bf Avg. Latency (s)                             & \bf Level 1 Avg. Lat. (s)                        & \bf Level 2 Avg. Lat. (s)                        & \bf Level 3 Avg. Lat. (s)                        \\
            \midrule
            Smolagent                    & 50.91         & 233.96                                           & 155.83                                           & 247.47                                           & 348.58                                           \\
            \, + SMAS (w/o Correction)   & 47.88         & 280.77 {\scriptsize \textcolor{red}{↑20.01\%}}   & 193.98 {\scriptsize \textcolor{red}{↑24.48\%}}   & 276.50 {\scriptsize \textcolor{red}{↑11.73\%}}   & 471.81 {\scriptsize \textcolor{red}{↑35.35\%}}   \\
            \, + SMAS (w/o Guidance)     & 48.48         & 271.95 {\scriptsize \textcolor{red}{↑16.24\%}}   & 211.04 {\scriptsize \textcolor{red}{↑35.43\%}}   & 291.22 {\scriptsize \textcolor{red}{↑17.68\%}}   & 332.38 {\scriptsize \textcolor{blue}{↓4.65\%}}   \\
            \, + SMAS (w/o Purification) & 49.70         & 236.21 {\scriptsize \textcolor{red}{↑0.96\%}}    & 137.00 {\scriptsize \textcolor{blue}{↓12.08\%}}  & 252.01 {\scriptsize \textcolor{red}{↑1.83\%}}    & 386.19 {\scriptsize \textcolor{red}{↑10.79\%}}   \\
            \, + SMAS                    & 50.91         & 321.15 {\scriptsize \textcolor{red}{↑37.27\%}}   & 233.11 {\scriptsize \textcolor{red}{↑49.59\%}}   & 320.93 {\scriptsize \textcolor{red}{↑29.68\%}}   & 501.31 {\scriptsize \textcolor{red}{↑43.81\%}}   \\
            \bottomrule
        \end{tabular}
    }
\end{table}

\subsubsection{Performance Analysis on Token-Intensive Scenarios}
\label{ablation_study_subset}

Complementing the comprehensive ablation study on the full GAIA benchmark (Table \ref{tab:ablation_full}), this section zooms in on the most demanding scenarios: the top-10 most token-intensive tasks per GAIA level. This analysis aims to evaluate the scalability and robustness of \method{} under extreme computational loads.

\paragraph{Amplified Efficiency and Robustness.}
As detailed in Table \ref{tab:ablation_subset}, the benefits of \method are significantly amplified in high-complexity regimes. While the average token reduction on the full dataset is 29.68\%, SMAS achieves a remarkable \textbf{50.13\%} reduction on this intensive subset. More importantly, unlike the full dataset where accuracy remains stable, SMAS yields a distinct accuracy improvement of \textbf{6.67\%} (from 40.00\% to 46.67\%) on these hard tasks. This suggests that as task complexity and context length increase, the Supervisor's interventions become indispensable not just for cost-saving, but for enabling the agent to complete tasks that were previously intractable due to context overflow or reasoning derailment.

\paragraph{Component Contribution in Extremes.}
The ablation results on this subset reinforce the synergistic roles of our three strategies. The \textit{w/o Purification} variant shows a drastic drop in efficiency (savings drop from 50.13\% to roughly 40\%), confirming that \textbf{Adaptive Observation Purification} is the primary countermeasure against the exponential token growth in complex tasks. Meanwhile, the removal of Correction or Guidance leads to a sharp decline in accuracy back to the baseline level (40.00\%), verifying that these modules are the key safety guardrails that allow the system to navigate long-horizon tasks successfully.

\begin{table}[!t]
    \caption{
        \textbf{Ablation study of \method's components} on the subset of GAIA benchmark (top-10 most token-intensive tasks per GAIA level).
    }
    \label{tab:ablation_subset}
    \centering
    \resizebox{\textwidth}{!}{
        \begin{tabular}{>{\raggedright\arraybackslash}m{4cm} c c c c c}
            \toprule
            \bf Method                   & \bf Avg. Acc.                                & \bf Avg. Token                                   & \bf Level 1 Avg. Token                           & \bf Level 2 Avg. Token                           & \bf Level 3 Avg. Token                           \\
            \midrule
            Smolagent                    & 40.00                                        & 1,446,526                                        & 933,013                                          & 2,037,437                                        & 1,369,131                                        \\
            \, + SMAS (w/o Correction)   & 40.00                                        & 719,075 {\scriptsize \textcolor{blue}{↓50.28\%}} & 426,786 {\scriptsize \textcolor{blue}{↓54.26\%}} & 755,543 {\scriptsize \textcolor{blue}{↓62.91\%}} & 974,895 {\scriptsize \textcolor{blue}{↓28.79\%}} \\
            \, + SMAS (w/o Guidance)     & 40.00                                        & 706,831 {\scriptsize \textcolor{blue}{↓51.14\%}} & 453,623 {\scriptsize \textcolor{blue}{↓51.38\%}} & 913,109 {\scriptsize \textcolor{blue}{↓55.18\%}} & 753,761 {\scriptsize \textcolor{blue}{↓44.95\%}} \\
            \, + SMAS (w/o Purification) & 46.67{\scriptsize \textcolor{red}{↑6.67\%}}  & 851,747 {\scriptsize \textcolor{blue}{↓41.11\%}} & 585,411 {\scriptsize \textcolor{blue}{↓37.26\%}} & 990,769 {\scriptsize \textcolor{blue}{↓51.37\%}} & 979,061 {\scriptsize \textcolor{blue}{↓28.49\%}} \\
            \, + SMAS                    & 46.67 {\scriptsize \textcolor{red}{↑6.67\%}} & 721,332 {\scriptsize \textcolor{blue}{↓50.13\%}} & 522,364 {\scriptsize \textcolor{blue}{↓44.01\%}} & 960,694 {\scriptsize \textcolor{blue}{↓52.85\%}} & 680,939 {\scriptsize \textcolor{blue}{↓50.26\%}} \\
            \bottomrule
        \end{tabular}
    }
\end{table}

\subsection{Failure Mode Analysis}

While \method demonstrated robustness across benchmarks, it relies on backbone LLMs and is thus subject to their inherent limitations. We identify three primary failure modes and their implications.

\paragraph{Information Loss during Purification.} The \textit{Adaptive Observation Purification} module faces a trade-off between context reduction and information preservation. In extreme cases observed in the OAgents framework, where single observations exceeded 200,000 characters, the Supervisor risks hallucinating or omitting critical details during compression. However, our empirical results suggest that the system's resilience to context overflow generally outweighs the cost of granular information loss, as evidenced by the overall token savings and success rates (see in Table \ref{tab:mas-agnostic}).

\paragraph{Ineffective Guidance and Loops.} The \textit{Inefficiency Guidance} module may occasionally provide suboptimal advice or fail to break a stubborn loop. To mitigate the risk of the Supervisor itself becoming a source of latency (e.g., engaging in an infinite correction loop with a non-responsive agent), we enforce a hard constraints of maximum \textit{two} guidance interventions per sub-task. While this design prioritizes bounded latency over guaranteed resolution for the hardest tasks, it effectively prevents runaway costs.

\paragraph{Variance in Trigger Frequency across Backbones.} Contrary to the assumption that a supervisor acts uniformly, our analysis reveals that \method exhibits different operational behaviors depending on the backbone model's capability. For instance, as shown in our logs with Qwen3-235B, less capable models trigger the \textit{Error Correction} module significantly more often due to frequent basic failures (e.g., malformed JSON tool calls). This frequent firing increases the Supervisor's token overhead, partially explaining the lower net token savings compared to GPT-4.1 (see in Figure \ref{fig:model-agnostic}). Conversely, stronger models may trigger supervision too rarely if the heuristic filter is not sensitive enough to their subtle logic errors. This highlights that while the \textit{framework} is model-agnostic, the \textit{efficiency gains} are correlated with the backbone LLM's adherence to instruction following.

\subsection{Case Study}
\label{case_study}

\begin{mdframed}[backgroundcolor=white!10,linecolor=black,linewidth=2pt,roundcorner=8pt]
    \textbf{\color{black!80}GAIA Benchmark Case Information}
    \vspace{5pt}
    \hrule
    \vspace{8pt}
    \textbf{Task ID}: 5b2a14e8-6e59-479c-80e3-4696e8980152 \\ \\
    \textbf{Level}: 3 \\ \\
    \textbf{Question}: The brand that makes these harnesses the dogs are wearing in the attached pic shares stories from their ambassadors on their website. What meat is mentioned in the story added Dec 8th 2022? \\ \\
    \textbf{Attached iamge}: 5b2a14e8-6e59-479c-80e3-4696e8980152.jpg
\end{mdframed}

\begin{tcolorbox}[
        colback=gray!5!white,
        colframe=gray!50!black,
        title=Smolagent,
        floatplacement=Hp,
        breakable  %
    ]
    \textbf{1. Activate first search\_agent}:\\
    \textbf{Failure attribution}: page\_down 10 times and end up searching \\
    \textbf{Output of first search\_agent}:
    \begin{lstlisting}[basicstyle=\ttfamily\small, breaklines=true, columns=fixed, breakindent=0pt, xleftmargin=0pt, gobble=0]
Observations: ### 1. Task outcome (short version):
Despite a detailed investigation and navigation through the RC Pets official website and blog section, I was unable to locate the ambassador story posted on December 8th, 2022. Therefore, I cannot report any mention of meat in its content.
\end{lstlisting}
    \textbf{2. Activate another search\_agent}:
    \begin{itemize}[leftmargin=1em]
        \item Plan: turn to access historical versions of the website
        \item page\_down for 7 times
        \item Then web\_search other resources
        \item Output of search\_agent
              \begin{lstlisting}[basicstyle=\ttfamily\small, breaklines=true, columns=fixed, breakindent=0pt, xleftmargin=0pt, gobble=0]
Observations: ### 1. Task outcome (short version):
Despite a detailed investigation and navigation through the RC Pets official website and blog section, I was unable to locate the ambassador story posted on December 8th,2022. Therefore, I cannot report any mention of meat in its content.
\end{lstlisting}
    \end{itemize}
    \textbf{3. Activate third search\_agent}:
    \begin{itemize}[leftmargin=1em]
        \item Plan: Thought: Since all direct website and archive attempts failed, I will extend my search to include RC Pets' official social media (Instagram, Facebook, Twitter) and broader web, as brands often share ambassador stories across platforms and tag or date posts.
        \item web\_search for 8 times
        \item Output of search\_agent
              \begin{lstlisting}[basicstyle=\ttfamily\small, breaklines=true, columns=fixed, breakindent=0pt, xleftmargin=0pt, gobble=0]
Observations: ### 1. Task outcome (short version):
No evidence was found of any RC Pets ambassador or partner story posted on December 8th, 2022, on official RC Pets Instagram, Facebook, Twitter, or the broader web. No blog posts, community group stories, or ambassador reposts regarding RC Pets for that date or month could be 
located, and no relevant content mentioning meat was found.
\end{lstlisting}
    \end{itemize}
    \textbf{Original final answer}:
    \begin{lstlisting}[basicstyle=\ttfamily\small, breaklines=true, columns=fixed, breakindent=0pt, xleftmargin=0pt, gobble=0]
Final answer: After an exhaustive investigation across all available RC Pets channels (official website, blog, and social media), Wayback Machine, pet influencer blogs, hashtags, and pet community forums, no ambassador story or repost by RC Pets or affiliated 
ambassadors/partners could be found for December 8th, 2022. No story mentioning any meat was identified for the date or month, and no content relevant to the request is accessible through any official or community web source. All logical avenues for retrieving the answer were 
attempted and exhausted.
\end{lstlisting}
    \textbf{Reformulated answer}: Unable to determine
\end{tcolorbox}

\begin{tcolorbox}[
        colback=gray!5!white,
        colframe=gray!50!black,
        title=Smolagent + SupervisorAgent,
        floatplacement=Hp,
        breakable  %
    ]
    \textbf{Supervision type: Basic\_extraction} - 8 times\\
    \textbf{Supervision type: Inefficiency\_analysis} - 2 times
    \begin{itemize}[leftmargin=1em]
        \item First time
              \begin{lstlisting}[basicstyle=\ttfamily\small, breaklines=true, columns=fixed, breakindent=0pt, xleftmargin=0pt, gobble=0]
guidance: Providing guidance to agent: 'Stop paging through the blog manually. Instead, use the web\_search tool or the Ruffwear website's internal search to find the specific ambassador story posted on December 8th, 2022. You could search for 'Ruffwear ambassador story December 8 2022' or check if the blog section has a date-filter or text search. Once you locate the precise article, examine it for any mention of meat, noting a quote for context.'
\end{lstlisting}
              \begin{lstlisting}[basicstyle=\ttfamily\small, breaklines=true, columns=fixed, breakindent=0pt, xleftmargin=0pt, gobble=0]
Analysis: The agent is stuck repeatedly using 'page\_down' without making meaningful progress toward locating the December 8th, 2022 story among the 'ambassadors' blog entries. There are 82 pages, and there's no indication the agent is close to the correct blog post. This paginated browsing is highly inefficient when a direct search or filter by date would be much faster. It's overwhelmingly more efficient to instead search for the specific blog post by date ('December 8th, 2022') using the web\_search tool or any site search functionality if available.
\end{lstlisting}
        \item Second time
              \begin{lstlisting}[basicstyle=\ttfamily\small, breaklines=true, columns=fixed, breakindent=0pt, xleftmargin=0pt, gobble=0]
approve
\end{lstlisting}
    \end{itemize}

    \textbf{Supervision type: sub\_agent\_result\_synthesis} - 1 time
    \begin{itemize}[leftmargin=1em]
        \item Original output length: 47902
        \item Length after process: 1438
              \begin{lstlisting}[basicstyle=\ttfamily\small, breaklines=true, columns=fixed, breakindent=0pt, xleftmargin=0pt, gobble=0]
content = """
### Official Ruffwear Ambassador Stories Section
- Ruffwear ambassador stories are shared in 'Explored: A Ruffwear Blog' (https://ruffwear.com/blogs/explored/tagged/ambassadors).

### Story Posted December 8, 2022
- Title: "Snow Camping With Theresa & Cassie"
- Author: Theresa Silveyra
- Date Published: December 8, 2022 ("2022-12-08 08:00:01 -0800")
- Story URL: https://ruffwear.com/blogs/explored/snow-camping-with-theresa-cassie

### Meat Mentioned in the Story
- The only specific meat mentioned is "bacon".

> Specific quoted context from the story:
> "Of course, barely a minute had passed before Cassie's longing eyes and dangling tongue signaled a desire to play or, perhaps, a desire for treats. She led the way back, her fluffy white tail raised high in excitement, ready to take on this new year and, more likely, ready to devour the New Year's Day bacon being cooked up in camp."

### Confirmation of Accuracy
- The article and details are confirmed by both page metadata and direct content search.
- Searches for "meat", "chicken", "beef", "turkey", "pork", "fish", "salami", and "sausage" yielded no other results; only bacon was present.

### Additional Context
- Author attribution and publication date are validated.
- Other foods or treats are referenced, but bacon is the only explicit meat in the text.

### Direct Answer
- The meat mentioned in the Ruffwear ambassador story posted on December 8th, 2022 is bacon.
"""
\end{lstlisting}
    \end{itemize}
\end{tcolorbox}

\subsection{Prompts}
\label{appendix:prompt}
\begin{tcolorbox}[
        colback=gray!5!white,
        colframe=gray!50!black,
        title=\textbf{Base prompt} for \method,
        floatplacement=Hp,
        breakable  %
    ]
    \begin{lstlisting}[basicstyle=\ttfamily\small, breaklines=true, columns=fixed, breakindent=0pt, xleftmargin=0pt, gobble=0]
base_prompt = f"""
Role: You are an expert supervisor in a multi-agent system. Your role is to monitor an agent's actions, ensure alignment with the main goal, correct errors, and optimize the workflow.

Objective: The overall objective (Global Task) is: "{global_task}"

Agent context:
You are currently reviewing an action from the agent '{agent_name}'.
This agent's specific sub-task (Local Task) is: '{local_task}'.
Here is the current local execution trace: {local_trace_str}
Here is the summary of the agent's latest thought process and the resulting observation:
    {summary}
    {current_state_summary}
    
Rules:
1.  Assess Necessity: First, assess if intervention is truly necessary. If the agent's action and observation are correct and productive, use the "approve" action. Avoid unnecessary interventions.
2.  Be Decisive: When an intervention is needed, choose the most effective action to move the project forward.
3.  Output Format: Your response MUST be a valid JSON object.

Actions:
Your available actions are: 
-  `approve`: The agent's action is correct and requires no changes. 
-  `correct_observation`: The observation contains errors or can be significantly improved (e.g., filtered, summarized, extracted). You will provide a corrected version. 
-  `provide_guidance`: The observation is correct, but the agent's thinking or next step is flawed. You will provide a hint or corrected reasoning to guide the agent. 
-  `run_verification`: You have doubts about the factual accuracy of the observation and need an external assistant to verify it. 

Your response MUST be a JSON object with the following structure:
{
  "analysis": "Your brief analysis of the situation, explaining your reasoning for the chosen action.",
  "action": "ONE of the available actions: ['approve', 'correct_observation', 'provide_guidance', 'run_verification']",
  "parameters": {
      "new_observation": "IF action is 'correct_observation', provide the refined observation here.",
      "guidance": "IF action is 'provide_guidance', provide a clear hint or instruction for the agent's next thought process.",
      "task": "IF action is 'run_verification', provide the verification question for the assistant."
  }
}"""

\end{lstlisting}
\end{tcolorbox}

\begin{tcolorbox}[
        colback=gray!5!white,
        colframe=gray!50!black,
        title=Prompt for "error\_occurrence",
        floatplacement=Hp,
        breakable  %
    ]
    \textbf{Base Prompt:}
    \begin{lstlisting}[basicstyle=\ttfamily\small, breaklines=true, columns=fixed, breakindent=0pt, xleftmargin=0pt, gobble=0]
...
\end{lstlisting}
    \textbf{Addtional Prompt:}
    \begin{lstlisting}[basicstyle=\ttfamily\small, breaklines=true, columns=fixed, breakindent=0pt, xleftmargin=0pt, gobble=0]
f"""
**Role**:You are an expert Debugger and AI Diagnostician. Your primary goal is to understand the root cause of an error and provide the most effective solution to get the agent back on track.

**Situation**:
The agent's last action resulted in a critical error, which is detailed in the "summary" of the agent's action below. **Approval is not an option; you must intervene.**

**--- Your Debugging Framework (MANDATORY) ---**
Before generating your JSON output, you MUST follow this structured thinking process:

**Step 1: Analyze the Error**
    - What is the precise error message and type (e.g., `Tool Error`, `Python Exception`, `APIError`)?

**Step 2: Examine the Context**
    - Review the `local_execution_trace` and the agent's `thought` process leading to the error.
    - What was the agent *trying* to accomplish?
    - Was the tool call or code it executed (`summary` section) syntactically correct but logically flawed?

 **Step 3: Root Cause Diagnosis**
    - Based on the error and context, what is the single most likely root cause?
    - (e.g., "The agent passed a natural language string to a tool expecting a mathematical expression.", "The agent is trying to access a file that does not exist.")

**Step 4: Formulate a Solution Strategy**
    - Based on the root cause, determine the best intervention:
        - If the error can be fixed by correcting the agent's **next thought process or action**, choose `provide_guidance`. This is the most common case for logical errors.
        - If the error was caused by faulty information in the **previous observation** that the agent is now acting upon, choose `correct_observation`.
        - If you lack critical information to solve the error and need to consult an external source, choose `run_verification`.

**--- YOUR ACTIONABLE OUTPUT (JSON) ---**
Based on your diagnosis, provide your final decision in the JSON format.
\end{lstlisting}
\end{tcolorbox}

\begin{tcolorbox}[
        colback=gray!5!white,
        colframe=gray!50!black,
        title=Prompt for "inefficiency\_behavior",
        floatplacement=Hp,
        breakable  %
    ]
    \textbf{Base Prompt:}
    \begin{lstlisting}[basicstyle=\ttfamily\small, breaklines=true, columns=fullflexible,xleftmargin=0pt]
...
\end{lstlisting}
    \textbf{Additional Prompt:}
    \begin{lstlisting}[basicstyle=\ttfamily\small, breaklines=true, columns=fixed, breakindent=0pt, xleftmargin=0pt, gobble=0]
f"""
**Role**:You are a pragmatic and experienced AI workflow strategist. Your primary goal is to ensure the agent team achieves its task in the most efficient way **from its current state**.

**Situation**:An inefficiency trigger has been activated for agent '{agent_name}'. **This is a flag for you to review, NOT a confirmation of a problem.** The agent might be engaged in a necessary, methodical process.

**Global Execution Trace**:
{global_trace_str}

**--- Your Decision Framework (MANDATORY) ---**
Before generating your JSON output, you MUST follow this structured thinking process:

**Step 1: Goal & Plan Inference**
    - Based on the `Global Execution Trace`, what is the agent's immediate, implicit plan?
    - (e.g., "The agent is clearly trying to collect all rows of a data table by repeatedly using `page_down`.")

**Step 2: Progress Assessment**
    - Is the agent making tangible progress towards its inferred goal?
    - Is each new step yielding new, relevant information (even if it's just more rows of the same table)?
    - How close is the agent to completing this sub-task? (e.g., "It is on page 10 of 13, it is very close to getting all the data.")

**Step 3: Cost-Benefit Analysis of Intervention**
    - **Compare two costs**:
        - **Cost A**: The estimated cost (time, tokens) of letting the agent **continue** its current, perhaps clumsy, path to completion.
        - **Cost B**: The estimated cost of **interrupting** the agent, guiding it to a new path, and having it **start over** on that new path.
        - **CRITICAL QUESTION**: Is the agent "one step away" from solving its sub-task? If so, interrupting it is almost always the wrong decision, even if a theoretically "better" path exists.

**Step 4: Decision and Justification**
    - Based on the analysis above, decide between `approve` and `provide_guidance`.
    
**--- YOUR ACTIONABLE OUTPUT (JSON) ---**
You must choose ONE of the following two actions:

**1. If you decide the agent should continue:**
    - **Condition**: The agent is making clear, incremental progress AND is close to completing its sub-task (Cost A < Cost B).
    - **Action**: MUST be `"approve"`.
    - **Analysis**: Briefly explain *why* the agent's current path, while perhaps repetitive, is the most pragmatic way forward from its current state. (e.g., "The agent is methodically paginating through a table to gather all data. Although repetitive, this is a valid and necessary process. It is on page 10 of 13 and about to succeed. Intervention would be disruptive.")

**2. If you decide the agent is truly stuck:**
    - **Condition**: The agent is in a non-productive loop (e.g., getting the same observation repeatedly) OR the alternative path is overwhelmingly more efficient and the agent is not close to finishing (Cost B << Cost A).
    - **Action**: MUST be `"provide_guidance"`.
    - **Analysis**: Briefly explain the root cause of the inefficiency.
    - **Guidance**: The `guidance` parameter MUST contain a clear, concrete, and actionable instruction that represents a *significantly* better strategy. (e.g., "Instead of scrolling, use the `web_search` tool with the query 'who had the most BB for the 1977 Yankees' to get the answer directly.")
"""
\end{lstlisting}
\end{tcolorbox}

\begin{tcolorbox}[
        colback=gray!5!white,
        colframe=gray!50!black,
        title=Prompt for "excessive observation length",
        floatplacement=Hp,
        breakable  %
    ]
    \begin{lstlisting}[basicstyle=\ttfamily\small, breaklines=true, columns=fixed, breakindent=0pt, xleftmargin=0pt, gobble=0]
f"""
# Role: AI Agent Observation Compressor
You are a specialized data compression model for an AI agent. Your sole purpose is to process raw observations (HTML, text, etc.) and reduce their token count while strictly preserving their structural integrity and all potentially useful information.

**## Core Principles ##**
1.  **Context-Agnostic:** You have NO knowledge of the agent's overall goal or past actions. Do NOT try to infer the task. Your compression must be generic and unbiased, preserving information that could be useful for ANY potential task.
2.  **Preservation Over Compression:** It is critically important to avoid over-summarization. Losing a potentially key piece of information is a greater failure than not compressing enough. The output must retain enough detail for the agent to make informed decisions.
3.  **Structural Integrity:** The output's structure (headings, lists, paragraphs, HTML hierarchy) must mirror the input's structure. Do not merge distinct sections.
4.  **Preserve Metadata**: Always keep leading lines like `"Address: ..."`, `"Viewport: ..."` verbatim. 

**##Compression Rules##**
Based on the type of content, apply the following rules:
    
### **Type 1: For HTML Content**
    Your goal is to simplify the HTML to its semantic and structural core, removing presentation-focused noise.
    1.  **Simplify Tags:** Remove non-essential attributes.
        -   **REMOVE attributes like:** `class`, `id`, `style`, `onclick`, `onmouseover`, and any `data-*` or `js-*` attributes. These are primarily for styling and scripting, not for content structure.
        -   **KEEP essential attributes:** `href`, `src`, `alt`, `title`, `aria-label`, `placeholder`, `value`. These attributes contain crucial information for navigation and interaction.
    2.  **Remove Non-Visible Content:** Completely remove `<script>`, `<style>`, and HTML comment `` blocks.
    3.  **Preserve Content:** Keep ALL text content within tags exactly as it is. Do not summarize the text inside the HTML.
    4.  **Whitespace:** Condense multiple spaces, newlines, and tabs in the HTML structure into a single space where appropriate to improve readability without losing structure.
    **Example:**
        * **Original:** `<td class='datacolBoxR' style='padding: 5px;'><a href="/wiki/some_link" title="Some Link">25</a></td>`
        * **Compressed:** `<td><a href="/wiki/some_link" title="Some Link">25</a></td>`

### **Type 2: For Plain Text Content**
    Your goal is to make the text more concise without losing factual information or its original layout.
    1.  **Retain Key Information:** Fully preserve all named entities (e.g., people, organizations, locations), numbers, dates, codes, IDs, and any factual data.
    2.  **Condense Prose:** For descriptive sentences or paragraphs, rephrase them to be more direct. Remove filler words, redundant phrases, and overly elaborate adjectives. However, do NOT eliminate the sentence entirely.
    3.  **Maintain Structure:** If the input text has multiple paragraphs, bullet points, or numbered lists, the output MUST have the same structure. Do not flatten a list into a single paragraph.
    **Example:**
        * **Original:** "The company, officially known as The International Business Machines Corporation (IBM), is a very large and influential American multinational technology corporation that has its headquarters located in Armonk, New York, and it was originally founded all the way back in 1911."
        * **Compressed:** "The International Business Machines Corporation (IBM) is an American multinational technology corporation headquartered in Armonk, New York, founded in 1911."

**## Final Instruction ##**
Process the following observation according to the rules above. Provide only the compressed output, without any extra text, explanation, or preamble.
    {observation}
"""
\end{lstlisting}
\end{tcolorbox}

\begin{tcolorbox}[
        colback=gray!5!white,
        colframe=gray!50!black,
        title=Prompt for "result\_synthesis",
        floatplacement=Hp,
        breakable  %
    ]
    \# this is for synthesis the final answer from sub-agent

    \textbf{Base Prompt:}
    \begin{lstlisting}[basicstyle=\ttfamily\small, breaklines=true, columns=fixed, breakindent=0pt, xleftmargin=0pt, gobble=0]
  ...
  \end{lstlisting}
    \textbf{Additional Prompt:}
    \begin{lstlisting}[basicstyle=\ttfamily\small, breaklines=true, columns=fixed, breakindent=0pt, xleftmargin=0pt, gobble=0]
  f"""
  **Role**:You are an expert Intelligence Analyst working for a manager agent. Your task is to process a verbose report from a sub-agent (e.g., a search specialist) and synthesize a direct, comprehensive, and clean answer for your manager.
  
  **--- YOUR INPUTS ---**
  **1. The Manager's Request (Immediate Goal)**:
      - "{local_task}"
  
  **2. The Overall Mission (Global Goal)**:
      - "{global_task}"
  
  **3. The Sub-Agent's Full Field Report (Raw Observation)**:
       ```
      {summary} 
      ```
  (Note: The 'summary' variable here contains the sub-agent's full, multi-part final_answer)
  
  **--- YOUR CRITICAL TASK ---**
  Your sole task is to read the ENTIRE "Field Report" (including the short version, detailed version, and the summary of work) and synthesize a single, clean, and self-contained response that **fully and completely** answers the "Manager's Request".
      **Critical Rule for Synthesis**:
      **Preserve Semantic Structure**: When synthesizing, you MUST maintain the original information's hierarchy. If the source contains headings, chapters, articles, or numbered/bulleted lists, these structural elements **MUST be preserved** in your output to give context to the data points below them. **Do not flatten a structured document into a simple, unstructured block of text.**
      **Your Internal Thought Process (MANDATORY)**:
          1.  **Deconstruct the Manager's Request**: What are the specific pieces of information the manager is asking for? Create a mental checklist.
          2.  **Scan the Entire Report**: Read all parts of the sub-agent's report to find the answers for your checklist. The most valuable details are often in the "extremely detailed version" or the "summary of work".
          3.  **Synthesize, Don't Just Extract**: Combine the findings into a coherent, fluent, and direct answer. Do not simply copy the "short version". Your answer must be comprehensive enough to prevent the manager from needing to ask follow-up questions.
          
  **Example**:
  - **Manager's Request**: "Find the number of encoder layers in the BERT-Base model."
  - **Sub-Agent's Report**: (A long text containing "Short version: 12 layers", "Detailed version: ...Section 3 of the paper states L=12 for BERT-Base...", etc.)
  - **Your Ideal Synthesized Output**: "The BERT-Base model has 12 encoder layers (L=12), as specified in Section 3 of the original paper by Devlin et al., 2018."
  
  **Action**:Your action MUST be `"correct_observation"`.
  
  **Parameter**:Provide your final, synthesized answer in the `"new_observation"` parameter.
  """
  \end{lstlisting}
\end{tcolorbox}

\end{document}